\documentclass[letterpaper,11pt]{article}

\pdfoutput=1

\usepackage{pgfplots}
\usepackage{multirow}
\usepackage{float}
\usepackage{amssymb}
\usepackage{dsfont}
\usepackage{mathrsfs}
\usepackage{enumitem}
\usepackage[T1]{fontenc}
\usepackage[british]{chet}
\usepackage[font=small,format=hang,labelfont={sf,bf}]{caption}
\usepackage{tikz}
\usepackage{float}
\usepackage{tensor}
\usepackage{mathtools}

\usetikzlibrary{calc, external, intersections, arrows.meta, arrows, patterns, plotmarks, shapes.geometric, pgfplots.groupplots}

\pgfplotsset{compat=1.18}
\usepgfplotslibrary{colorbrewer}

\makeatletter
\g@addto@macro\bfseries{\boldmath}
\makeatother

\DeclareMathOperator{\Rea}{Re}
\DeclareMathOperator{\Ima}{Im}
\DeclareMathOperator{\Tr}{Tr}

\newcommand{\lsp}{\hspace{1pt}}
\newcommand{\llsp}{\hspace{0.5pt}}
\newcommand{\lnsp}{\hspace{-1pt}}

\renewcommand{\leq}{\leqslant}
\newcommand{\nc}[3]{\lsp\tensor[^{#2}]{#1}{_{#3}}}

\allowdisplaybreaks

\definecolor{forestgreen}{rgb}{0.0, 0.27, 0.13}
\definecolor{darkblue}{rgb}{0.0, 0.0, 0.55}
\definecolor{darkred}{rgb}{0.55, 0.0, 0.0}
\hypersetup{colorlinks,
           linkcolor={forestgreen},
           citecolor={darkblue},
           urlcolor={darkred},
           pdftitle={Gradient RG Flow in Scalar-Fermion QFTs},
           pdfdisplaydoctitle
}

\tikzset{cross/.style={path picture={
      \draw[black]
            (path picture bounding box.south east) --
            (path picture bounding box.north west)
            (path picture bounding box.south west) --
            (path picture bounding box.north east);}}}

\title{Gradient RG Flow in Scalar-Fermion QFTs}

\author{William H.\ Pannell, William P.\ Ronayne, and Andreas Stergiou\emails{(\href{mailto:william.pannell@kcl.ac.uk}{william.pannell}, \href{mailto:william.ronayne@kcl.ac.uk}{william.ronayne}, \href{mailto:andreas.stergiou@kcl.ac.uk}{andreas.stergiou})@kcl.ac.uk}}

\affiliation{Department of Mathematics, King's College London, Strand, London WC2R 2LS, United Kingdom}

\abstract{The gradient property of the renormalisation group (RG) is examined to four-loop order in scalar-fermion systems in $d=4$ and $d=4-\varepsilon$ dimensions. The crucial role played by the beta shift, which is a modification of the standard dim-reg beta function, is elucidated, and specific conditions that it needs to satisfy for the RG flow to be gradient are derived. Over a thousand gradient-flow conditions are found, all of which are scheme-independent and satisfied whenever the full set of results needed to check them is available. It is shown, in the framework of the $\varepsilon=4-d$ expansion, that the space of conformal field theories (CFTs) is dominated by those with non-zero beta shift as the number of fields grows. Physical properties of CFTs obtained as solutions where the beta functions are not zero in the $\varepsilon$ expansion are discussed.}

\date{November 2025}

\begin{document}

\maketitle

\toc

\section{Introduction}
Perturbation theory provides essential hints for the structure of quantum field theory (QFT). In the context of the renormalisation group (RG), this has been vividly elucidated by the $\varepsilon$ expansion \cite{Wilson:1971dc}, which captures features of strongly coupled physics despite its perturbative starting point. Perturbative considerations have also suggested that the RG flow has a gradient property, starting with \cite{Wallace:1974dx, Wallace:1974dy}. This has recently been extended to high loop order for multi-scalar systems in $d=4$ and $d=4-\varepsilon$~\cite{Jack:2018oec, Pannell:2024sia, Pannell:2025ixz}, as  well as to multi-scalar systems in $d=6$ and $d=6-\varepsilon$ \cite{Gracey:2015fia, Benfatto:2025yci}. In this work we provide further evidence for the gradient property of the RG by showing that it holds to four-loop order in scalar-fermion systems in $d=4$ and $d=4-\varepsilon$.

Essential to the structure of the gradient RG flow is the role of a contribution to the trace of the energy-momentum tensor which is present whenever there exist non-conserved spin-one operators, $J^\mu_A$, of dimension close to $d-1$, where the subscript $A$ labels a collection of flavour indices. Operator mixing effects imply that
\begin{equation}
    T^\mu{\!}_\mu=\beta^I\mathcal{O}_I+S^A\partial_\mu J^\mu_A\,,
\end{equation}
where all operators are fully renormalised, the index $I$ labels again a collection of flavour indices associated with the local operators $\mathcal{O}_I$ that are responsible for the RG flow, and $\beta^I$ is the beta function associated with their couplings $g^I$. Use of the equations of motion, which are always available in a perturbative Lagrangian setting, allows the rewriting
\begin{equation}
    T^\mu{\!}_\mu=\big(\beta^I-(Sg)^I\big)\mathcal{O}_I\,,
\end{equation}
where the meaning of $(Sg)^I$ will become clear in examples below. This was first understood in \cite{Jack:1990eb}. The essential point is that the proper vector field in coupling space that generates the RG flow is not the traditional beta function $\beta^I$ but
\begin{equation}
    B^I=\beta^I-(Sg)^I\,.
\end{equation}
Moreover, fixed points should be found as points in coupling space where $B^I=0$ as opposed to $\beta^I=0$. It was proven in \cite{Fortin:2012hn} that if $\beta^I=0$ in a unitary theory then $(Sg)^I=0$ too, but there are also solutions to $B^I=0$ that do not require both contributions to vanish separately. Such solutions correspond to CFTs in which the couplings flow in limit cycles of the beta function vector field \cite{Fortin:2011ks, Fortin:2011sz}. The leading contribution to $S^A$ in scalar-fermion systems, where it appears at three loops, was computed in \cite{Fortin:2012hn} for the first time, and later confirmed in \cite{Jack:2013sha} and extended in \cite{Davies:2021mnc} to theories with gauge fields in $d=4$ and $d=4-\varepsilon$. We will refer to $S^A$ as the ``beta shift'' below.

A solution with $S^A\neq0$ was first found some time ago in $d=4-\varepsilon$ in \cite{Fortin:2011ks}, where it was initially misinterpreted as corresponding to a theory with scale but not conformal invariance \cite{Fortin:2011sz}. This was later understood to be a point with $\beta^I\neq0$ but $B^I=0$ \cite{Fortin:2012hn}, thus describing a conformal field theory. Despite the fact that such solutions are known to exist, though, it is not clear how common they are in the $\varepsilon$ expansion below $d=4$ dimensions. While an example in a theory with only scalar fields is currently lacking, we find in this work that such fixed points are ubiquitous in scalar-fermion systems. In fact, we find that as we increase the number of fields in our system, fixed points with non-zero $S^A$ dominate the space of solutions to $B^I=0$. The inclusion of gauge fields allows perturbative fixed points to be found in $d=4$ too, and such a fixed point with non-zero beta shift has also been found \cite{Fortin:2012cq}.

Since the leading-order contributions to $S^A$ appear at high loop order, namely five for purely scalar and three for scalar-fermion systems in $d=4-\varepsilon$, one always starts by seeking roots of the one-loop beta functions, for $\beta^I$ and $B^I$ are equal at one loop. As we see in this work, any such root that leads to non-zero $S^A$ will cease to be a root of $\beta^I$ at the $\varepsilon$ order at which the $S^A$ contribution is non-zero. It will, however, survive as a root of $B^I\ne\beta^I$ at that loop order. This is a reflection of the absence of scale without conformal invariance in these systems \cite{Fortin:2012hn}. Indeed, if a root $g_*^I$ with non-zero $S^A_*$ remained a root of $\beta^I$, i.e.\ $\beta^I(g_*)=0$, then $T^\mu{\!}_\mu$ would be equal to $S_*^A\partial_\mu J^\mu_A$, which is the condition for a conserved dilatation current \cite{Polchinski:1987dy}.

Just as in its role for the absence of scale without conformal invariance, the beta shift is also crucial for the gradient flow of the RG. The question here is whether a Riemannian metric, $G_{IJ}$, and scalar quantity, $A$, both functions of the couplings $g^I$, exist such that
\begin{equation}\label{eq:gradflow}
    \frac{\partial A}{\partial g^I}=G_{IJ}B^J\,.
\end{equation}
It is known that an equation like this for $\beta^I$ fails to hold at six loops in multi-scalar systems \cite{Pannell:2024sia}, while it continues to hold for $B^I$, subject to some constraints on $S^A$. These constraints have been explicitly derived in \cite{Pannell:2024sia}, but remain unverified to date due to the lack of a six-loop determination of the beta shift in multi-scalar models. Besides these constraints on the beta shift there are numerous others, involving highly non-trivial combinations of beta function coefficients, that are satisfied. Therefore we are led to the belief that the constraints on the beta shift are also satisfied and the flow described by $B^I$ is indeed gradient in multi-scalar systems at six loops. It is worth noting that if we drop the condition that the metric be Riemannian, then an equation like \eqref{eq:gradflow} holds to all orders in perturbation theory \cite{Jack:1990eb, Osborn:1991gm}.

The gradient flow equation, \eqref{eq:gradflow}, provides a generalisation of Zamolodchikov's $c$-theorem \cite{Zamolodchikov:1986gt} to higher dimensions, where $A$ takes the role of $c$ as the RG monotone. Generalisations of the $c$-theorem to higher dimensions have long been sought, and while there has been some success in proving weak monotonicity theorems \cite{Myers:2010xs,Jafferis:2011zi,Klebanov:2011gs,Casini:2012ei,Komargodski:2011vj,Hartman:2023qdn}, considerably less is known about the gradient property, the strongest version of monotonicity one can demand. As it appears that proper handling of the beta shift is crucial for this property, the lack of examples of scalar fixed points with non-zero beta shift means that studying purely multiscalar gradient flow will be insufficient to truly illuminate any deeper structure. We must turn instead to systems which can house fixed points with non-zero beta shift, i.e.\ scalar-fermion theories.

The complication compared to the multiscalar case considered previously is the presence of two distinct types of couplings, quartic and Yukawa, which means that we must divide indices into a doublet $(I,\alpha)$, where $\lambda_{ijkl}\to\lambda_I$ and $y_{iab}\to y_\alpha$. The gradient flow equation is thus most naturally written as the vector equation
\begin{equation}
    \begin{pmatrix}
        \partial_I A \\\partial_\alpha A
    \end{pmatrix}=\begin{pmatrix}
        G_{IJ} & G_{I\beta}\\ G_{\alpha J} & G_{\alpha\beta}
    \end{pmatrix}\begin{pmatrix}
        \beta^J\\ \beta^\beta
    \end{pmatrix}.
    \label{eq:GradFloweqn}
\end{equation}
It is crucial to note, as will become apparent later when examining the constraints on the coefficients in the beta function which arise from this system of equations, that we cannot simultaneously unit-normalise the Kronecker delta part of $G_{IJ}$ and $G_{\alpha\beta}$, as was first noticed in \cite{Fortin:2012ic}. In order for the purely scalar solution previously considered in \cite{Pannell:2024sia} to be embedded as a subsector of our solution, we will choose the normalisation
\begin{equation}
    G_{IJ}=\delta_{IJ}+\text{O}(\lambda,y^2)\,,\qquad\qquad G_{\alpha\beta}=\nc{g}{0}{}\lsp\delta_{\alpha\beta}+\text{O}(\lambda,y^2)\,,
\end{equation}
for $\nc{g}{0}{}$ which will be fixed as part of the solution. We will then seek to make sense of \eqref{eq:GradFloweqn} in perturbation theory, where we can expand $A$, $G$ and $\beta$ as a series the couplings, and truncate at a certain loop order to produce a finite system of equations. By solving these equations explicitly, we will be able to construct an RG monotone, $A$, and Riemannian metric, $G_{IJ}$, through four loops.

The structure of this paper is as follows. In Section \ref{sec:fp} we provide analytic examples of fixed points at which the beta shift does not vanish, and demonstrate that at these fixed points the solution to $\beta=0$ ceases to exist at high enough loop order. In Section \ref{sec:gradflowsec} we construct a solution to gradient flow, in $d=4$ and also in $d=4-\varepsilon$ where there exist RG flows between non-trivial fixed points. We find that the existence of a solution is equivalent to 1016 scheme-invariant constraints on the coefficients appearing in the beta function, and crucially that these constraints can only be satisfied once the beta shift has been taken into account. We then detail how this solution provides an extension of the conjectured $\widetilde{F}$-theorem \cite{Fei:2015oha, Fei:2016sgs} to gradient flow. Finally, in section \ref{sec:numerics} we numerically trawl the space of fixed points for low numbers of scalars and fermions, finding that fixed points with non-zero beta shift not only exist, but become increasingly frequent for larger numbers of fields. We conclude in Section \ref{sec:conc}. An ancillary \href{https://www.wolfram.com/mathematica/}{\texttt{Mathematica}} file attached to the \href{https://arxiv.org}{\texttt{arXiv}} submission of this paper can be used by the interested reader to inspect the beta functions, beta shifts and gradient constraints we have obtained in $d=4$ and $d=4-\varepsilon$.

\section{Fixed points}\label{sec:fp}
In this section we study fixed points of scalar-fermion theories in $4-\varepsilon$ dimensions which have non-zero beta shift. We will consider theories with general quartic and Yukawa interactions for $N_s$ real scalars and $N_f$ Weyl fermions. The most general interaction Lagrangian is
\begin{equation}\label{eq:Lint}
    \mathscr{L}_{\text{int}} = \tfrac{1}{4!}\lambda_{ijkl}\phi_i\phi_j\phi_k\phi_l + (\tfrac12 y_{iab}\phi_i\psi_a\psi_b + \text{h.c.})\,,
\end{equation}
where $i,j,k,l=1,\ldots,N_s$, $a,b=1,\ldots,N_f$, $\lambda_{ijkl}$ is real and fully symmetric, and $y_{iab}$ is complex and symmetric in $a$ and $b$. General expressions for the beta functions of such theories have been calculated in terms of the tensors $\lambda_{ijkl}$ and $y_{iab}$ \cite{Jack:1982hf, Jack:1983sk, Jack:1984vj, Machacek:1983tz, Machacek:1983fi, Machacek:1984zw, Steudtner:2024teg, Steudtner:2025blh}. The one loop beta functions are presented in Appendix \ref{app:betasinvs}. The two and three loop beta functions are lengthy expressions and we will not reproduce them here---they are presented in a clear form in \cite{Jack:2024sjr}. The highest loop order results known to date are in \cite{Steudtner:2025blh}.

The flavour symmetry group of \eqref{eq:Lint} is $O(N_s)\times U(N_f)$. This is a symmetry of the kinetic term of the Lagrangian in the traditional sense, and a symmetry of \eqref{eq:Lint} if we  take the couplings $\lambda_{ijkl}$ and $y_{iab}$ to transform as tensors under $O(N_S)\times U(N_f)$, i.e.\ the indices $i, j, k, l$ transform under $O(N_S)$ and the indices $a, b$ under $U(N_f)$.

We have $S=S^AT_A$, where $T_A$ runs through the Lie algebra generators of $O(N_s)\times U(N_f)$. Consequently, the beta shift is a direct sum of a real antisymmetric matrix $S_{ij}$ which acts on the scalar flavour indices and a complex anti-Hermitian matrix $P_{ab}$ which acts on the fermionic flavour indices. The form of the beta shift is simply given by the infinitesimal transformation of the coupling tensors under $O(N_s)\times U(N_f)$,
\begin{equation}\label{eq:Saction}
    \begin{aligned}
        (S\lambda)_{ijkl} &= S_{ii'}\lambda_{i'jkl} + S_{jj'}\lambda_{ij'kl} + S_{kk'}\lambda_{ijk'l} + S_{ll'}\lambda_{ijkl'}\,,  \\
        (Sy)_{iab} &= S_{ii'}y_{i'ab} + P_{aa'}y_{ia'b} +y_{iab'}P_{b'b}\,.
    \end{aligned}
\end{equation}

Expressions for the beta shift have been calculated for the theories considered here \cite{Fortin:2012hn, Steudtner:2024teg}. The leading contribution is found at three loops. The scalar contribution is given by
\begin{equation}\label{eq:Sthreeloops}
    S_{ij}^{(1)}=\tfrac58\Tr(y_iy^*{\hspace{-5pt}}_ky_ly^*{\hspace{-5pt}}_m)\lambda_{jklm} + \tfrac38\Tr(y_iy^*{\hspace{-5pt}}_ky_ly^*{\hspace{-5pt}}_ly_jy^*{\hspace{-5pt}}_k) + \text{h.c.}-\{ i\leftrightarrow j\}\,,
\end{equation}
where the trace is over the fermionic flavour indices and we have rescaled $\lambda\to 16\pi^2\lambda$ and $y\to4\pi y$. The fermionic part $P_{ab}$ was zero for all the fixed points we studied and so we will not discuss it further. An expression for it can be found in \cite{Jack:2024sjr}, where it is referred to as $v_\psi$.

To solve the beta function equations in the $\varepsilon$ expansion, we first need to expand out the tensorial expressions explicitly into components, yielding an equation for each coupling. We do this using the \href{https://www.nikhef.nl/~form/}{\texttt{FORM}} language. This can also be done with the \href{https://www.wolfram.com/mathematica/}{\texttt{Mathematica}} package \href{https://github.com/aethomsen/RGBeta}{\texttt{RGBeta}} \cite{Thomsen:2021ncy}. Once the explicit equations are generated, they can be imported into \href{https://www.wolfram.com/mathematica/}{\texttt{Mathematica}} and solved. In $d=4-\varepsilon$, zeroes of the beta functions can be found order by order in an $\varepsilon$ expansion of the couplings. In general, the loop order corresponds to the order in $\varepsilon$, so only the one-loop beta function terms contribute to the values of the critical couplings at leading order, the two loop terms to the next order, and so on. The equations at leading order will involve non-linear polynomials with many roots, however all equations at higher order are linear. The leading contribution of the beta shift is at order $\varepsilon^3$, and this contribution is determined solely by the one loop roots of the beta functions. Correspondingly in the $B$ function, the beta shift will only contribute at third loop order and higher. Therefore it is necessary to go to three loop order to study the effects of the beta shift.

\subsection{Fixed points with non-zero beta shift}
\subsubsection{Two scalars and one Weyl fermion}
The simplest example of a fixed point with non-zero beta shift is found in a theory with two real scalars and one Weyl fermion. The interaction Lagrangian \eqref{eq:Lint} becomes
\begin{equation}\label{2s1f_lag}
    \mathscr{L}_{\text{int}}= \tfrac{1}{24}\lambda_1\phi_1^4 + \tfrac{1}{6}\lambda_2\phi_1^3\phi_2+\tfrac{1}{4}\lambda_3\phi_1^2\phi_2^2 + \tfrac{1}{6}\lambda_4\phi_1\phi_2^3 + \tfrac{1}{24}\lambda_5\phi_2^4 + (\tfrac12 y_1\phi_1\psi^2 + \tfrac12y_2\phi_2\psi^2 + \text{h.c.})\,.
\end{equation}
This theory, which has a total of nine real couplings, was also studied in \cite{Fortin:2011ks}. It is straightforward to solve the beta function equations analytically in \href{https://www.wolfram.com/mathematica/}{\texttt{Mathematica}}. The leading order equations for the Yukawa couplings are independent of the quartic couplings and so we solve these first. When solving the equations analytically, one finds a large number of solutions, including continuous families of solutions. This is expected as the solutions to the one loop equations are invariant under $O(N_s)\times U(N_f)$ rotations of the couplings \cite{Pannell:2023tzc, Osborn:2017ucf}. Consequently, many of these solutions represent the same fixed point. For a given solution of the Yukawa equations, there will then be a number of corresponding solutions for the scalar couplings.

One simple solution of the equations with non-zero beta shift is given by
\begin{equation}\label{eq:sol1expl}
     \begin{gathered}
         \lambda_{1}^{(1)} = \tfrac{821326 - 5427\sqrt{419802}}{9725376}\,,\qquad
         \lambda_{2}^{(1)} = \tfrac{7 \sqrt{\tfrac{469}{74}\,(3601 + 6\sqrt{419802})}}{131424}\,, \\
         \lambda_{3}^{(1)} = \tfrac{469(222 + \sqrt{419802})}{9725376}\,,\qquad
         \lambda_{4}^{(1)} = \tfrac{67 \sqrt{\tfrac{469}{74}(3601 + 6\sqrt{419802})}}{131424}\,,\qquad
         \lambda_{5}^{(1)} = \tfrac{7(373922 - 141 \sqrt{419802})}{9725376}\,,\\
         y_{1}^{(1)}= 0\,,\qquad
         y_{2}^{(1)} = \tfrac{1}{2\sqrt{2}}\,,
     \end{gathered}
\end{equation}
after the rescalings $\lambda\to 16\pi^2\lambda$ and $y\to4\pi y$. Since there are two scalars, the beta shift is an antisymmetric $2\times 2$ matrix, which is determined by
\begin{equation}
    S_{12}^{(1)}=\tfrac{35 \sqrt{\tfrac{469}{74}(3601 + 6 \sqrt{419802})}}{33644544}\,.
\end{equation}
This is obtained from the three-loop result \eqref{eq:Sthreeloops} upon using \eqref{eq:sol1expl}. The roots at second order can also be easily found using the beta function, but at order $\varepsilon^3$ the $B$ function must be used. The beta function for the imaginary part of $y_2$  at this order is equal to a non-zero constant, and so our solution to $\beta=0$ cannot be continued beyond two loops. If however we use the $B$ function, we find that the beta shift contribution precisely cancels this constant term, setting the equation identically zero and rendering the equations solvable at three loops. The breakdown of the perturbative solution of the beta function equations is a corollary of the fact that theories which possess scale but not conformal invariance are impossible perturbatively in unitary theories, as discussed in the introduction.

\subsubsection{Two scalars and two Weyl fermions}
Another relatively simple example of a fixed point with non-zero beta shift is given by considering two real scalars and two Weyl fermions. The interaction Lagrangian in this case is given by
\begin{equation}\label{eq:Lagtwotwo}
\begin{aligned}
    \mathscr{L}_{\text{int}}&= \tfrac{1}{24}\lambda_1\phi_1^4 + \tfrac{1}{6}\lambda_2\phi_1^3\phi_2+\tfrac{1}{4}\lambda_3\phi_1^2\phi_2^2 + \tfrac{1}{6}\lambda_4\phi_1\phi_2^3 + \tfrac{1}{24}\lambda_5\phi_2^4 \\
    &+ (\tfrac12y_1\phi_1\psi_1^2 + \tfrac12 y_2\phi_2\psi_1^2 + y_3\phi_1\psi_1\psi_2 + y_4\phi_2\psi_1\psi_2 + \tfrac12 y_5 \phi_1\psi_2^2 + \tfrac12 y_6\phi_2\psi_2^2+ \text{h.c.})\,.
\end{aligned}
\end{equation}
At one loop we solve for zeroes of the beta functions just as before. The solution at leading order is given by
\begin{equation}\label{eq:sol2expl}
\begin{gathered}
        \lambda^{(1)}_{1} = \tfrac{7087 + 357 \sqrt{52953}}{205770}\,, \qquad
        \lambda^{(1)}_{2} = -\tfrac{2 \sqrt{\tfrac{17}{19}(757 - 3 \sqrt{52953})}}{5415}\,,\\
        \lambda^{(1)}_{3} = \tfrac{17 (57 - \sqrt{52953})}{102885}\,,\qquad
        \lambda^{(1)}_{4} = -\tfrac{17 \sqrt{\tfrac{17}{19}(757 - 3 \sqrt{52953})}}{5415}\,,\qquad
        \lambda^{(1)}_{5} = \tfrac{4 (6346 + 9 \sqrt{52953})}{102885}\,,\\
        y^{(1)}_{1} =-y^{(1)}_{5} = \tfrac{1}{\sqrt{10}}\,,\qquad
        y^{(1)}_{2}=y^{(1)}_{3} =y^{(1)}_{4}=y^{(1)}_{6}= 0\,,
    \end{gathered}
\end{equation}
after the rescalings $\lambda\to 16\pi^2\lambda$ and $y\to4\pi y$. The corresponding leading contribution to the beta shift for this solution is determined by \eqref{eq:Sthreeloops} to be
\begin{equation}
    S_{12}^{(1)}=-\tfrac{\sqrt{\tfrac{17}{19}(757 - 3 \sqrt{52953})}}{108300}\,.
\end{equation}
These roots of the beta functions get extended to order $\varepsilon^2$, but at order $\varepsilon^3$ there are incompatible equations that arise from the beta functions of $y_2$ and $y_6$, which require $y_{2}^{(3)}+y_{6}^{(3)}$ to be equal to both plus and minus a non-zero constant. After including the contributions to the full $B$ function the incompatibility is removed and the solution \eqref{eq:sol2expl} can be extended to order $\varepsilon^3$.

\subsection{Physical interpretation and anomalous dimensions}
In the previous section, we studied two fixed points which are conformal but have non-zero beta functions. A question then arises as to the meaning of a fixed point where the couplings are apparently scale dependent. Typically, the Callan--Symanzik equation for the two-point function of fundamental fields, $\Phi$, can be used to show that the eigenvectors of the anomalous dimension matrix $\gamma =Z^{-1} \mu\frac{d}{d\mu}Z$ have definite scaling dimensions, with the corresponding anomalous dimensions given by the eigenvalues of $\gamma$. (Here we use $\Phi_0=Z\Phi$ as the relation between bare and renormalised fields.) When the beta function is not zero, this calculation does not work. This can be amended by using a different basis of fields, as we now show.

Let us first revisit the RG flow of the couplings in these CFTs \cite{Fortin:2011sz, Fortin:2012hn}. At a fixed point where $B_I=0$ and $\beta_I\neq 0$, we have
\begin{equation} \label{eq:betadiff}
    \mu\frac{d}{d\mu}g_I=(Sg)_I\,.
\end{equation}
It was shown in \cite{Fortin:2012hn} that $S$ is constant when $B_I=0$. Thus, the above is simply a linear differential equation for the vector of couplings $g$ given by a constant matrix $S$. We can integrate it to obtain
\begin{equation}
    g(\mu)=R(U^{-1})g(\mu_0)\,,\qquad U=\exp \left(-S\log\frac{\mu}{\mu_0}\right)\,,
\end{equation}
where the meaning of $R(U)$ becomes clear from \eqref{eq:Saction}, i.e.\ $R(U)$ is a product of exponentials of $S$. Since $S$ is in the Lie algebra of $O(N_s)\times U(N_f)$, we see the couplings undergo flavour rotations along the RG flow, and so the theory corresponds to a cycle of the RG flow.

Now, the theory is invariant under simultaneous flavour rotations of the fields and couplings. Consider the transformed quantities
\begin{equation}
    \bar{g} = R(U)g\,,\qquad \bar{\Phi} = U \Phi\,,\qquad \bar{Z} = Z U^{-1}\,.
\end{equation}
It is clear from \eqref{eq:betadiff} that $\mu \frac{d}{d\mu}\bar{g}=0$. Since $\mu\frac{d}{d\mu}(\bar{Z}\bar{\Phi})=0$, we find
\begin{equation} \label{eq:gammabar}
    \bar{\gamma}=U(\gamma+S)U^{-1}\,,
\end{equation}
where $\mu\frac{d}{d\mu}\bar{\Phi}=-\bar{\gamma}\bar{\Phi}$. In terms of the fields and couplings $\bar{\Phi}, \bar{g}$, the theory is now a standard CFT and the anomalous dimensions can be calculated from $\bar{\gamma}$ in the usual way. The conclusion of the above discussion is that in theories with non-zero beta shift, the physical anomalous dimension matrix is given by \eqref{eq:gammabar}. In particular, the eigenvalues are obtained from $\Gamma=\gamma+S$, which was identified in previous work as the matrix that is invariant under an ambiguity in choosing the field strength renormalisation factors \cite{Jack:1990eb, Fortin:2012hn, Herren:2021yur}.

It is important to note that while the matrix $\gamma$ can be taken to be symmetric, the matrix $\Gamma$ cannot. Nevertheless, eigenvalues of $\Gamma$ must be real at unitary fixed points. In the $\varepsilon$ expansion this typically occurs due to the fact that $S$ receives contributions at higher order than $\gamma$.

\section{Gradient flow}\label{sec:gradflowsec}
Let us now turn to the gradient flow equation, (\ref{eq:GradFloweqn}), which we will aim to solve order by order in perturbation theory. Each object in play will have an expansion in terms of tensor structures, which we can organise as
\begin{equation}
\begin{split}
    A&=\sum_n\sum_m\nc{a}{n}{m}\nc{A}{n}{m}\,, \\
    G_{\circ\star}&=\sum_n\sum_m\nc{g}{n}{m}(\nc{G}{n}{m})_{\circ\star}\,,\\
    \beta^\star&=\sum_n\sum_m\nc{\mathfrak{b}}{n}{m}(\nc{\beta}{n}{m})^\star\,,
\end{split}
\label{eq:gradplayers}
\end{equation}
where $\circ,\,\star\in(I,\alpha)$ represent arbitrary indices, $n$ counts the number of tensors in the structure, given by $n=N_\lambda+N_y/2$, and $m$ indexes a sum over the different possible tensor structures with the same value of $n$. Note that this notation agrees with that used in \cite{Pannell:2025ixz}, but differs from that used in \cite{Pannell:2024sia} where $n$ instead referred to the loop order at which the coefficient would contribute to the gradient flow equation. The coefficients $\nc{\mathfrak{b}}{n}{m}$ have been calculated in the $\overline{\text{MS}}$ renormalisation scheme through four loops in \cite{Steudtner:2024teg,Steudtner:2025blh}, though one should note that we will typically leave these coefficients unfixed and plug in the $\overline{\text{MS}}$ values only to verify that constraint equations on the beta function are satisfied. It is important to note that the values of $n$ over which the sum ranges can differ depending upon the indices one chooses for $\circ$ and $\star$, for instance with the sum in $\beta^I$ ranging over $\mathbb{Z}_{>0}$ while the sum in $\beta^\alpha$ ranges over $\mathbb{Z}_{>0} +\frac12$.

This counting of $n$ lays bare another complication compared to the multiscalar case: loop order mixing. Consider a structure $\nc{A}{n}{m}$ which contains both $\lambda$'s and $y$'s. Removing a $\lambda$, by the action of $\partial_I$, will cause a contribution to some $(\nc{\beta}{n-1}{m})^I$ in (\ref{eq:GradFloweqn}) which appears at $n-2$ loops. Removing a $y$, on the other hand, will contribute to some $(\nc{\beta}{n-\frac{1}{2}}{m})^\alpha$, which appears at $n-1$ loops. One thus sees that it is natural to associate $n$-loop diagrams in $\beta_\lambda$ with $n+1$ loop diagrams in $\beta_y$, as the gradient flow equation will mix their coefficients. Similarly, each of the off-diagonal $G_{I\alpha}$ terms in (\ref{eq:GradFloweqn}) will contribute to (\ref{eq:GradFloweqn}) at two different orders: first, e.g., at $n$ loops with $G_{I\beta}\beta^\beta$ and then again at $n+1$ loops with $G_{\alpha J}\beta^J$.

It is here that we should note a subtlety with the definition of the interaction tensor $y_{iab}$. As we begin near four dimensions, the actions considered in this paper are constructed using four-dimensional spinor representations, Weyl and Dirac, but the fixed points will correspond to non-trivial CFTs in three dimensions with an $\varepsilon\rightarrow1$ limit. At the end of this limit, the four-dimensional representations will all become reducible, with the 4d Weyl represenation splitting into two 3d Majorana fermions. As explained in \cite{Jack:2024sjr,Pannell:2023tzc}, it is possible to implement this splitting in an ad-hoc manner by writing the $N_f$ Weyl fermions, $\psi_a$ in terms of $N=2N_f$ Majorana fermions, $\{\zeta_a,\xi_a\}$, as
\begin{equation}
    \psi_a=\zeta_a+i\xi_a\,.
\end{equation}
With this definition, we can rewrite the Yukawa interaction as
\begin{equation}
\phi_i(y_{iab}\psi_a\psi_b+y^*_{iab}\bar{\psi}_a\bar{\psi}_b)=
\phi_i\big(2\Rea(y_{iab})(\zeta_a\zeta_b-\xi_a\xi_b)+2\Ima(y_{iab})(\zeta_a\xi_b+\zeta_b\xi_a)\big)=g_{iAB}\phi_i\theta_A\theta_B\,,
\label{eq:weyltomajorana}
\end{equation}
where $A,B=1,\ldots,2N_f$, $\zeta_a=\theta_a$ and $\xi_a=\theta_{N_f+a}$. One thus sees that it is possible to realise all of the fixed points using manifestly real Yukawa tensors. It is considerably easier to construct the graphs using these real tensors, given that the fermion lines will not be directional, and one must thus be mindful of the fact that the $N$ of this section is twice the number of fermions $N_f$ appearing in other sections, and that strict correspondence requires using the map (\ref{eq:weyltomajorana}). As \cite{Jack:2024sjr,Steudtner:2025blh} also use real Yukawa tensors to write down the beta function through four loops, taking $y$ to be real will make it easier to directly use their results for the $\nc{\mathfrak{b}}{n}{m}$. Though we will not do so in this paper, one can solve the beta function for $g_{iAB}$ not of the form compatible with (\ref{eq:weyltomajorana}), and thus construct 3d fixed points which have no realisation in terms of 4d representations. Such fixed points were studied in \cite{Pannell:2023tzc}.

Using (\ref{eq:weyltomajorana}), one sees that any trace containing an odd number of the real couplings $g_{i AB}$ must vanish if it is to descend from a four-dimensional action written in terms of Weyl fermions. However, in principle such terms may appear in the beta function for genuinely three-dimensional theories, and spoil the uplift to a four-dimensional theory. This difficulty was recognised in \cite{Zerf:2017zqi,Steudtner:2025blh}, where it was found that at four-loops such terms, proportional to the trace over five $g$'s, can begin to appear in the beta function with non-zero coefficients. Using consistency conditions from three dimensional supersymmetric theories, it was noted in \cite{Steudtner:2025blh} that the coefficients of such terms can be computed consistently, to produce a general beta function capable of capturing fixed points both with and without an uplift to four-dimensions. To be fully general, we will include diagrams with odd traces, and simply set the coefficients of such diagrams in the beta function to zero as necessary. Note that in the remainder of the section we will use $y$, rather than $g$ to refer to the real Yukawa tensors, and drop the distinction between the indices $A$ and $a$.

Solving (\ref{eq:GradFloweqn}) in full generality through a given loop order requires that one first write down all of the tensors $\nc{A}{n}{m}$, $(\nc{G}{n}{m})_{\circ\star}$ and $(\nc{\beta}{n}{m})^\star$ which appear on the right-hand side of (\ref{eq:gradplayers}). However, one notices that by contracting the free indices in $(\nc{G}{n}{m})_{\circ\star}$ or $(\nc{\beta}{n}{m})^\star$ with the appropriate interaction vertex, one obtains a tensor structure with no external indices. It is thus sufficient to write down only the list of vacuum diagrams, $\nc{A}{n}{m}$, from which we then produce $(\nc{G}{n}{m})_{\circ\star}$ and $(\nc{\beta}{n}{m})^\star$ by removing all possible combinations of two vertices or one vertex, respectively. As we only wish to capture contributions to $\beta^\star$ which are connected, we will only accept vacuum diagrams which are one-vertex irreducible.

To produce the list of vacuum diagrams, we first note that it is possible, by cutting edges, to split any vacuum diagram into a pair of, potentially disconnected, graphs, one of which contains only $y^\alpha$ vertices and another which contains only $\lambda^I$ vertices. It is then straightforward to construct the list of purely Yukawa and purely scalar graphs iteratively, attaching an extra vertex in every possible way to all graphs of one order lower, and removing identical graphs using graph isomorphism algorithms. For this we use two \href{https://www.wolfram.com/mathematica/}{\texttt{Mathematica}} packages, the \href{http://www.xact.es/}{\texttt{xAct}} with \href{http://www.xact.es/xtras/}{\texttt{xTras}} package \cite{NUTMA20141719} to construct the tensors $y_{iab}$ and $\lambda_{ijkl}$, and then the \href{http://szhorvat.net/pelican/igraphm-a-mathematica-interface-for-igraph.html}{\texttt{IGraph/M}} package\cite{Horvát2023} to implement the graph isomorphisms. As a technical point, in order for the graph algorithms to properly distinguish between the fermion and scalar lines, it is crucial that we work with coloured graphs, colouring each edge based on whether it is a scalar index $i$ or a fermionic index $a$.

As we wish to test gradient flow through $3/4$ loops, we will need to construct all of the vacuum graphs through $\text{O}(\lambda^{5},\,\lambda^4y^2,\,\ldots,\,y^{10})$, and in the end find the number of graphs given in Table \ref{tab:graphnum}. One should note that this method of constructing diagrams produces terms which contain non-overlapping divergences and thus trivially vanish in $\overline{\text{MS}}$. As we are interested in constructing the gradient flow equations in a generic scheme, and will not want to restrict to $\overline{\text{MS}}$ from the start, it is crucial to include these diagrams, which generically can have non-zero coefficient. Removing these diagrams, as well as the diagrams with an odd number of $y$'s appearing in traces discussed above, produces precisely the graphs identified previously in \cite{Jack:2024sjr}.
\begin{table}[H]
    \centering
    \begin{tabular}{|c|c|c|c|c|}
    \hline
        Loop order in $\beta_\lambda$/$\beta_y$: & 0/1 & 1/2 & 2/3 & 3/4  \\
        \hline
         Terms in $A$ & 4 & 12 & 53 & 309 \\
         Terms in $G_{\circ\star}$ & -- & 8 & 59 & 593 \\
         Terms in $\beta^\star$ & 4 & 22 & 168 & 1739 \\
         \hline
    \end{tabular}
    \caption{The total number of graphs contributing to the gradient flow equation at each loop order we consider. Note that for $0/1$ loop order only the $\delta_{\circ\star}$ part of the metric will contribute to the gradient flow equation.}
    \label{tab:graphnum}
\end{table}
One can see immediately that the number of equations will be greater than the number of unknowns already at $1/2$ loop order, so that we find constraints on the beta function at next-to-leading order. This is to be contrasted with the multiscalar case where the number of equations grows more slowly, and constraints begin to appear only at four loops.

\subsection{Solution in four dimensions}
After enumerating all of the graphs, solving (\ref{eq:GradFloweqn}) proceeds by evaluating the contractions order by order, and setting the coefficients of the same tensor structures on opposite sides of the equation equal to each other. To illustrate this process, let us examine the equations explicitly through 1/2 loop order. The $A$-function will be
\begin{equation}\label{eq:Adef}
\begin{split}
    A(\lambda,y)=&\nc{a}{2}{1}\,
+\text{O}(\varepsilon^4)\,,
\end{split}
\end{equation}
where each small filled circle represents a $\lambda$-vertex, each small square represents a $y$-vertex, and scalar lines are shown as dashed. The corresponding $\beta$-functions will be
\begin{equation}
    \beta^I=\nc{\mathfrak{b}}{2}{1}\,%
+\text{O}(\varepsilon^{\frac{7}{2}})\,,
\end{split}
\end{equation}
where there is implicitly a sum over the distinct permutations of external indices. Finally, the three submatrices in the metric will be
\begin{equation}
    G_{IJ}=%
\,+\,\text{O}(\varepsilon^2)\,,
\end{split}
\end{equation}
where this is again an implicit symmetrisation over the external indices in the metric. Note that we use different conventions for the metric and the beta function: for the beta function there is just a sum over permutations, but for the metric there is an additional symmetry factor normalising the expression.

At leading order, only the trivial term will contribute from the metric, so that the gradient flow equations can be straightforwardly computed just by acting with a derivative on $A$:
\begin{equation}\label{eq:01loopsol}
\begin{split}
    2\nc{a}{2}{1}=0\,,\qquad 4\nc{a}{2}{2}=\nc{g}{0}{}\nc{\mathfrak{b}}{\frac{3}{2}}{1}\,,\\ 4\nc{a}{2}{3}=\nc{g}{0}{}\nc{\mathfrak{b}}{\frac{3}{2}}{2}\,,\qquad 2\nc{a}{2}{4}=\nc{g}{0}{}\nc{\mathfrak{b}}{\frac{3}{2}}{3}\,,
\end{split}
\end{equation}
where the last equation has a factor of 2 rather than a factor of 4 because of the two distinct permutations of external indices. These solutions can clearly be solved using just the $\nc{a}{2}{m}$, leaving $\nc{g}{0}{}$ as a free parameter. At the next order, the scalar part of the gradient flow equations is again straightforward to write down, as the metric will contribute trivially:
\begin{equation}
    \nc{a}{3}{1}=\nc{\mathfrak{b}}{2}{1}\,,\qquad \nc{a}{3}{2}=3\nc{\mathfrak{b}}{2}{2}\,,\qquad\nc{a}{3}{3}=2\nc{\mathfrak{b}}{2}{3}\,.
\end{equation}

For the fermion part of the equations, we must now take into account the non-trivial nature of the metric. Suppose that we wish to determine the equation associated with a specific graph $\gamma$ in $\beta^\star$, which is $\text{O}(\varepsilon^n)$. Any contribution to this equation from $G_{\circ\star}\beta^\star$ will schematically take the form
\begin{equation}
\begin{tikzpicture}[baseline=(vert_cent.base)]
        \node (vert_cent) at (0,0) {$\phantom{\cdot}$};
        \draw (1.5,0.2) -- (0,0.4);
        \draw (1.5,-0.2) -- (0,-0.4);
        \draw (0,0.2) -- (-1,0.5);
        \draw (0,-0.2) -- (-1,-0.5);
        \draw[fill=white] (1.5,0) circle (0.25cm);
        \draw[pattern=crosshatch, pattern color=black] (1.5,0) circle (0.25cm);
        \draw[fill=white] (0,0) circle (0.5cm);
        \draw[pattern=north west lines, pattern color=black] (0,0) circle (0.5cm);
        \node[yshift=4pt,scale=1.5] at (-0.75,0) {$\vdots$};
        \node[yshift=3.5pt,scale=1.25] at (0.875,0) {$\vdots$};
        \node[yshift=7pt] at (0,0.5) {$G_{\circ \star}$};
        \node[yshift=7pt] at (1.5,0.25) {$\beta^\star$};
    \end{tikzpicture}\,,
\end{equation}
where because we want to be generic we have not specified the index, or distinguished between scalar and fermion lines. One sees that this simply highlights a specific subdivergent graph, $\gamma'$, in the graph $\gamma$, so that to obtain the total of all contributions from the metric, one simply must sum over all of the possible subdiverences, akin to a forest structure. That is, for each diagram $\gamma$ the right-hand side of the gradient flow equation will then take the form
\begin{equation}\label{eq:gradflowrightside}
    \sum_{\gamma'\subseteq\gamma}S_{\gamma',\gamma}(\nc{g}{n-m}{\gamma-\gamma'})\nc{\mathfrak{b}}{m}{\gamma'}\,,
\end{equation}
where $m=N^{\gamma'}_\lambda+N^{\gamma'}_y/2$ counts the number of vertices in the subgraph $\gamma'$, and the $S_{\gamma',\gamma}$ are symmetry factors. By $\gamma-\gamma'$, we mean the graph one obtains by completely removing all of the vertices in $\gamma'$ from the graph $\gamma$, but retaining as external lines any edges connecting $\gamma'$ to the rest of the graph. If $M_{\gamma'}$ is the number of ways of producing $\gamma'$ from the contraction of the metric diagram $\gamma-\gamma'$ with $\gamma'$, and $P_\gamma$ is the number of distinct permutations of the external indices in $\gamma$, then the symmetry factors are given by
\begin{equation}
    S_{\gamma',\gamma}=\frac{M_{\gamma'}}{P_\gamma}\,.
\end{equation}
For instance, consider the graph
\begin{equation}
    \gamma=\begin{tikzpicture}[scale=0.5,baseline=(vert_cent.base),square/.style={regular polygon,regular polygon sides=4}]
    \node (vert_cent) at (0,0) {$\phantom{\cdot}$};
    \def\radius{0.8cm}
    \draw[thick] (0,0) circle[radius=\radius];
    \node[inner sep=0pt,outer sep=0pt,square, draw,fill=white,scale=3pt] (l) at (180:\radius) {};
    \node[inner sep=0pt,outer sep=0pt,square, draw,fill=white,scale=3pt] (r) at (0:\radius) {};
    \node[inner sep=0pt,outer sep=0pt,square, draw,fill=white,scale=3pt] (t) at (90:\radius) {};
    \node[inner sep=0pt,outer sep=0pt,square, draw,fill=white,scale=3pt] (b) at (270:\radius) {};
    \draw[dashed] (t) to (b);
    \draw[dashed] (l) to (-1.5,0);
    \draw[densely dashed] (r) to (1.5,0);
    \draw[thick] (1.5,0) to (2,1);
    \draw[thick] (1.5,0) to (2,-1);
    \node[inner sep=0pt,outer sep=0pt,square,draw,fill=white,scale=3pt] at (1.5,0) {};
    \draw[color=red] (-1.1,-1) rectangle (0.4,1);
    \draw[color=red] (1.1,-1.2) rectangle (-0.4,1.2);
    \draw[color=red] (-1.3,-1.4) rectangle (1.75,1.4);
    \end{tikzpicture}
\end{equation}
with three divergent subgraphs (including the graph itself) marked by the red rectangles. Excising these rectangles will produce three distinct contributions:
    \begin{equation}
    \begin{tikzpicture}[scale=0.5,baseline=(vert_cent.base),square/.style={regular polygon,regular polygon sides=4}]
    \node (vert_cent) at (0,0) {$\phantom{\cdot}$};
    \def\radius{0.8cm}
    \draw[thick] (0,0) circle[radius=\radius];
    \node[inner sep=0pt,outer sep=0pt,square, draw,fill=white,scale=3pt] (l) at (180:\radius) {};
    \node[inner sep=0pt,outer sep=0pt,square, draw,fill=white,scale=3pt] (r) at (0:\radius) {};
    \node[inner sep=0pt,outer sep=0pt,square, draw,fill=white,scale=3pt] (t) at (90:\radius) {};
    \node[inner sep=0pt,outer sep=0pt,square, draw,fill=white,scale=3pt] (b) at (270:\radius) {};
    \draw[dashed] (t) to (b);
    \draw[dashed] (l) to (-1.5,0);
    \draw[densely dashed] (r) to (1.5,0);
    \draw[thick] (1.5,0) to (2,1);
    \draw[thick] (1.5,0) to (2,-1);
    \node[inner sep=0pt,outer sep=0pt,square,draw,fill=white,scale=3pt] at (1.5,0) {};
    \draw[color=red] (-1.1,-1) rectangle (0.4,1);
    \draw[color=red] (1.1,-1.2) rectangle (-0.4,1.2);
    \draw[color=red] (-1.3,-1.4) rectangle (1.75,1.4);
    \end{tikzpicture}\longrightarrow S_1\,\begin{tikzpicture}[scale=0.65,
    square/.style={regular polygon,regular polygon sides=4},baseline=(vert_cent.base)]
    \node (vert_cent) at (0,-1) {$\phantom{\cdot}$};
    \foreach [count=\i] \coord in {
(1,-0.5), (-1,-0.5)}{
        \node[inner sep=0pt,outer sep=0pt] (p\i) at \coord {};
    }
    \foreach [count=\i] \coord in {
(-1,-1), (1,-1),(-1,-1.5),(1,-1.5)}{
        \node[inner sep=0pt,outer sep=0pt] (d\i) at \coord {};
    }
    \node[inner sep=0pt,outer sep=0pt,square, draw,fill=white,scale=3pt] (r) at (0.25,-1.25) {};
    \node[inner sep=0pt,outer sep=0pt,square, draw,fill=white,scale=3pt] (l) at (-0.25,-1.25) {};
    \draw[thick] (1.5,-1) arc (90:-90:0.25cm);
    \draw[densely dashed] (1.5,-1) arc (90:270:0.25cm);
    \foreach [count=\i] \coord in {
(1.5,-1.5), (1.5,-1),(1.75,-1.25)}{
        \node[inner sep=0pt,outer sep=0pt,square,draw,fill=white,scale=3pt] (b\i) at \coord {};
    }
    \draw[thick] (d1) to (l);
    \draw[thick] (d3) to (l);
    \draw[densely dashed] (r) to (l);
    \draw[thick] (r) to (d2);
    \draw[thick] (r) to (d4);
    \draw[densely dashed] (p1) edge (p2);
    \draw[thick] (d4) to (b1);
    \draw[thick] (d2) to (b2);
    \draw[densely dashed] (b3) to[bend right=60] (p1);
    \draw[color=red] (1,-1.7) rectangle (2,-0.3);
    \node at (0,0.2) {$\nc{g}{1}{1}$};
    \node at (1.5,0.3) {$\nc{\mathfrak{b}}{\frac{3}{2}}{2}$};
\end{tikzpicture}+S_2\,\begin{tikzpicture}[scale=0.65,
    square/.style={regular polygon,regular polygon sides=4},baseline=(vert_cent.base)]
    \node (vert_cent) at (0,-1) {$\phantom{\cdot}$};
    \foreach [count=\i] \coord in {
(1,-0.5), (-1,-0.5)}{
        \node[inner sep=0pt,outer sep=0pt] (p\i) at \coord {};
    }
    \foreach [count=\i] \coord in {
(-1,-1), (1,-1),(-1,-1.5),(1,-1.5)}{
        \node[inner sep=0pt,outer sep=0pt] (d\i) at \coord {};
    }
    \node[inner sep=0pt,outer sep=0pt,square, draw,fill=white,scale=3pt] (t) at (0,-0.75) {};
    \node[inner sep=0pt,outer sep=0pt,square, draw,fill=white,scale=3pt] (b) at (0,-1.25) {};
    \draw[thick] (1.5,-1) arc (90:-90:0.25cm);
    \draw[densely dashed] (1.5,-1) arc (90:270:0.25cm);
    \foreach [count=\i] \coord in {
(1.5,-1.5), (1.5,-1),(1.75,-1.25)}{
        \node[inner sep=0pt,outer sep=0pt,square,draw,fill=white,scale=3pt] (b\i) at \coord {};
    }
    \draw[thick] (d1) to (b);
    \draw[thick] (d3) to (b);
    \draw[thick] (t) to (d2);
    \draw[thick] (t) to (d4);
    \draw[densely dashed] (p1) to (b);
    \draw[densely dashed] (p2) to (t);
    \draw[thick] (d4) to (b1);
    \draw[thick] (d2) to (b2);
    \draw[densely dashed] (b3) to[bend right=60] (p1);
    \draw[color=red] (1,-1.7) rectangle (2,-0.3);
    \node at (0,0.2) {$\nc{g}{1}{2}$};
    \node at (1.5,0.3) {$\nc{\mathfrak{b}}{\frac{3}{2}}{2}$};
\end{tikzpicture}+S_3\,\begin{tikzpicture}[scale=0.5,baseline=(vert_cent.base),square/.style={regular polygon,regular polygon sides=4}]
    \node (vert_cent) at (0,-0.5) {$\phantom{\cdot}$};
    \def\radius{0.8cm}
    \draw[thick] (0,0) circle[radius=\radius];
    \node[inner sep=0pt,outer sep=0pt,square, draw,fill=white,scale=3pt] (l) at (180:\radius) {};
    \node[inner sep=0pt,outer sep=0pt,square, draw,fill=white,scale=3pt] (r) at (0:\radius) {};
    \node[inner sep=0pt,outer sep=0pt,square, draw,fill=white,scale=3pt] (t) at (90:\radius) {};
    \node[inner sep=0pt,outer sep=0pt,square, draw,fill=white,scale=3pt] (b) at (270:\radius) {};
    \draw[dashed] (t) to (b);
    \draw[densely dashed] (l) to (-2.5,0);
    \draw[densely dashed] (r) to[bend left=60] (0,-1.2);
    \draw[thick] (0,-1.2) to (-1.2,-0.5);
    \draw[thick] (0,-1.2) to (-1.2,-1);
    \draw[thick] (-1.2,-0.5) to (-2.5,-0.5);
    \draw[thick] (-1.2,-1) to (-2.5,-1);
    \node[inner sep=0pt,outer sep=0pt,square,draw,fill=white,scale=3pt] at (0,-1.2) {};
    \draw[color=red] (-1.2,-1.4) rectangle (1.2,1);
    \node at (-1.85,1.6) {$\nc{g}{0}{}$};
    \node at (0,1.7) {$\nc{\mathfrak{b}}{\frac{5}{2}}{16}$};
    \end{tikzpicture}\,.
\end{equation}
In this case, as $\gamma$ is symmetric on its external fermionic legs its symmetry factor is $P_\gamma=1$. Similarly, as the sub-diagrams in the metric and in $\beta^\alpha$ are manifestly symmetric on their fermionic indices, $M_{\gamma'}=1$ in each case, so that $S_1=S_2=S_3=1$.

In principle, one can then write down the gradient flow equations by exhaustively applying the above technique to each diagram in $\beta^\star$ at each loop order. In practice, however, it is easier to produce the gradient flow equations by brute force on a computer by simply computing all of the possible contractions of each diagram in the metric with each diagram in the beta function, separating out the resulting graphs into isomorphism classes after the fact. In the end, one finds that the fermion part of the gradient flow equations are
\begin{gather}
    (\nc{g}{0}{}) \nc{\mathfrak{b}}{\frac{5}{2}}{1}-4\nc{a}{3}{2}=0\,,\qquad (\nc{g}{0}{}) \nc{\mathfrak{b}}{\frac{5}{2}}{2}-2 \nc{a}{3}{3}=0\,,\qquad -\nc{a}{3}{4}+(\nc{g}{0}{}) \nc{\mathfrak{b}}{\frac{5}{2}}{3}+\frac{\nc{g}{1}{7} \nc{\mathfrak{b}}{\frac{3}{2}}{1}}{2}+\frac{\nc{g}{1}{8} \nc{\mathfrak{b}}{\frac{3}{2}}{1}}{2}=0\,,\nonumber\\ -\nc{a}{3}{4}+(\nc{g}{0}{})
   \nc{\mathfrak{b}}{\frac{5}{2}}{4}+\frac{\nc{g}{1}{6} \nc{\mathfrak{b}}{\frac{3}{2}}{1}}{2}+\nc{g}{1}{3} \nc{\mathfrak{b}}{\frac{3}{2}}{3}=0\,,\qquad -2 \nc{a}{3}{4}+(\nc{g}{0}{}) \nc{\mathfrak{b}}{\frac{5}{2}}{5}+2 \nc{g}{1}{1} \nc{\mathfrak{b}}{\frac{3}{2}}{3}+2 \nc{g}{1}{2} \nc{\mathfrak{b}}{\frac{3}{2}}{3}=0\,,\nonumber\\ -2 \nc{a}{3}{5}+(\nc{g}{0}{})
   \nc{\mathfrak{b}}{\frac{5}{2}}{6}+\frac{\nc{g}{1}{7} \nc{\mathfrak{b}}{\frac{3}{2}}{3}}{2}+\frac{\nc{g}{1}{8} \nc{\mathfrak{b}}{\frac{3}{2}}{3}}{2}=0\,,\qquad -2 \nc{a}{3}{5}+(\nc{g}{0}{}) \nc{\mathfrak{b}}{\frac{5}{2}}{7}+\nc{g}{1}{6} \nc{\mathfrak{b}}{\frac{3}{2}}{3}=0\,,\nonumber\\ (\nc{g}{0}{}) \nc{\mathfrak{b}}{\frac{5}{2}}{8}-6 \nc{a}{3}{6}=0\,,\qquad -\nc{a}{3}{7}+(\nc{g}{0}{})
   \nc{\mathfrak{b}}{\frac{5}{2}}{9}+\frac{\nc{g}{1}{7} \nc{\mathfrak{b}}{\frac{3}{2}}{2}}{2}+\frac{\nc{g}{1}{8} \nc{\mathfrak{b}}{\frac{3}{2}}{2}}{2}=0\,,\nonumber\\ -\nc{a}{3}{7}+(\nc{g}{0}{}) \nc{\mathfrak{b}}{\frac{5}{2}}{10}+\frac{\nc{g}{1}{5} \nc{\mathfrak{b}}{\frac{3}{2}}{3}}{2}+\frac{\nc{g}{1}{6} \nc{\mathfrak{b}}{\frac{3}{2}}{2}}{2}=0\,,\qquad -\nc{a}{3}{7}+(\nc{g}{0}{})
   \nc{\mathfrak{b}}{\frac{5}{2}}{11}+\frac{\nc{g}{1}{5} \nc{\mathfrak{b}}{\frac{3}{2}}{3}}{2}+\nc{g}{1}{4} \nc{\mathfrak{b}}{\frac{3}{2}}{3}=0\,,\nonumber\\ (\nc{g}{0}{}) \nc{\mathfrak{b}}{\frac{5}{2}}{12}-6 \nc{a}{3}{8}=0\,,\qquad -2 \nc{a}{3}{9}+(\nc{g}{0}{}) \nc{\mathfrak{b}}{\frac{5}{2}}{13}+\frac{\nc{g}{1}{5} \nc{\mathfrak{b}}{\frac{3}{2}}{2}}{2}=0\,,\qquad -2 \nc{a}{3}{9}+(\nc{g}{0}{})
   \nc{\mathfrak{b}}{\frac{5}{2}}{14}+\nc{g}{1}{4} \nc{\mathfrak{b}}{\frac{3}{2}}{2}=0\,,\nonumber\\ -2 \nc{a}{3}{10}+(\nc{g}{0}{}) \nc{\mathfrak{b}}{\frac{5}{2}}{15}+\nc{g}{1}{3} \nc{\mathfrak{b}}{\frac{3}{2}}{2}+\nc{g}{1}{4} \nc{\mathfrak{b}}{\frac{3}{2}}{1}=0\,,\qquad -2 \nc{a}{3}{10}+(\nc{g}{0}{}) \nc{\mathfrak{b}}{\frac{5}{2}}{16}+\nc{g}{1}{1} \nc{\mathfrak{b}}{\frac{3}{2}}{2}+\nc{g}{1}{2}
   \nc{\mathfrak{b}}{\frac{3}{2}}{2}=0\,,\nonumber\\ -2 \nc{a}{3}{10}+(\nc{g}{0}{}) \nc{\mathfrak{b}}{\frac{5}{2}}{17}+\nc{g}{1}{5} \nc{\mathfrak{b}}{\frac{3}{2}}{1}=0\,,\qquad -6 \nc{a}{3}{11}+(\nc{g}{0}{}) \nc{\mathfrak{b}}{\frac{5}{2}}{18}+\nc{g}{1}{1} \nc{\mathfrak{b}}{\frac{3}{2}}{1}+\nc{g}{1}{2} \nc{\mathfrak{b}}{\frac{3}{2}}{1}+\nc{g}{1}{3} \nc{\mathfrak{b}}{\frac{3}{2}}{1}=0\,,\nonumber\\ -3
   \nc{a}{3}{12}+(\nc{g}{0}{}) \nc{\mathfrak{b}}{\frac{5}{2}}{19}+\frac{\nc{g}{1}{6} \nc{\mathfrak{b}}{\frac{3}{2}}{3}}{2}+\frac{\nc{g}{1}{7} \nc{\mathfrak{b}}{\frac{3}{2}}{3}}{2}+\frac{\nc{g}{1}{8} \nc{\mathfrak{b}}{\frac{3}{2}}{3}}{2}=0\,.\label{eq:eqns12}
\end{gather}

Up until this point we have not specified the values of the $\nc{\mathfrak{b}}{n}{m}$, and one can attempt to solve these equations as they are using the unfixed coefficients $\nc{a}{n}{m}$ and $\nc{g}{n}{m}$. However, one finds that these equations generically don't admit a solution, and that after solving for as many of the coefficients in $A$ and in $G_{IJ}$ as one can, one is left with four equations written purely in terms of the $\nc{\mathfrak{b}}{n}{m}$:
\begin{equation}\label{eq:12constraints}
\begin{gathered}
    3\nc{\mathfrak{b}}{2}{2}\nc{\mathfrak{b}}{\frac{5}{2}}{2}-\nc{\mathfrak{b}}{2}{3}\nc{\mathfrak{b}}{\frac{5}{2}}{1}=0\,, \\
    2\nc{\mathfrak{b}}{\frac{3}{2}}{1}(\nc{\mathfrak{b}}{\frac{5}{2}}{10}-\nc{\mathfrak{b}}{\frac{5}{2}}{11})+\nc{\mathfrak{b}}{\frac{3}{2}}{2}(-2\nc{\mathfrak{b}}{\frac{5}{2}}{4}+\nc{\mathfrak{b}}{\frac{5}{2}}{5})+2\nc{\mathfrak{b}}{\frac{3}{2}}{3}(\nc{\mathfrak{b}}{\frac{5}{2}}{15}-\nc{\mathfrak{b}}{\frac{5}{2}}{16})=0\,,\\(\nc{\mathfrak{b}}{\frac{3}{2}}{3})^2(\nc{\mathfrak{b}}{\frac{5}{2}}{13}-\nc{\mathfrak{b}}{\frac{5}{2}}{14})-\nc{\mathfrak{b}}{\frac{3}{2}}{3}\nc{\mathfrak{b}}{\frac{3}{2}}{2}(\nc{\mathfrak{b}}{\frac{5}{2}}{9}-2\nc{\mathfrak{b}}{\frac{5}{2}}{10}+\nc{\mathfrak{b}}{\frac{5}{2}}{11})+(\nc{\mathfrak{b}}{\frac{3}{2}}{2})^2(\nc{\mathfrak{b}}{\frac{5}{2}}{6}-\nc{\mathfrak{b}}{\frac{5}{2}}{7})=0\,,  \\
    \nc{\mathfrak{b}}{\frac{3}{2}}{2}\nc{\mathfrak{b}}{\frac{3}{2}}{3}(\nc{\mathfrak{b}}{\frac{5}{2}}{3}-\nc{\mathfrak{b}}{\frac{5}{2}}{4})+\nc{\mathfrak{b}}{\frac{3}{2}}{2}\nc{\mathfrak{b}}{\frac{3}{2}}{1}(\nc{\mathfrak{b}}{\frac{5}{2}}{6}-\nc{\mathfrak{b}}{\frac{5}{2}}{7})-\nc{\mathfrak{b}}{\frac{3}{2}}{1}\nc{\mathfrak{b}}{\frac{3}{2}}{3}(2\nc{\mathfrak{b}}{\frac{5}{2}}{9}-3\nc{\mathfrak{b}}{\frac{5}{2}}{10}+\nc{\mathfrak{b}}{\frac{5}{2}}{11})\\
    \hspace{10cm}+(\nc{\mathfrak{b}}{\frac{3}{2}}{3})^2
(\nc{\mathfrak{b}}{\frac{5}{2}}{15}-\nc{\mathfrak{b}}{\frac{5}{2}}{17})=0\,.\end{gathered}
\end{equation}
Note that these equations are equivalent to the four constraints found in \cite[Eq.\ (9.4)]{Jack:2024sjr}. These are constraint equations which the components of $\beta^\star$ must obey if it is to be a gradient vector field. The necessity of constraints at this loop order can be seen already from Table \ref{tab:graphnum}. As there will be 22 equations which we must solve with only 20 unknowns, we should expect at least two constraint equations. Plugging in the $\overline{\text{MS}}$ values for the $\nc{\mathfrak{b}}{n}{m}$, taken from \cite{Jack:2024sjr}, one finds that these equations are all satisfied, and thus that the beta function is gradient at this loop order. As the beta shift $S^A$ appears only beginning at three loops, it is not necessary to consider a difference between $\beta^\star$ and $B^\star$ at this order.

The first of these constraints can be given a simple interpretation, and in fact belongs to a family of constraints which will appear at all higher loop orders as well. To demonstrate how this family of constraints is constructed, let us examine the four beta function diagrams which are involved in more detail, and construct explicitly the right-hand side of the gradient flow equation for these graphs, given by (\ref{eq:gradflowrightside}). The diagrams in question are
\begin{equation}
    \begin{tikzpicture}[scale=0.5,baseline=(vert_cent.base),square/.style={regular polygon,regular polygon sides=4}]
        \node (vert_cent) at (0,0) {$\phantom{\cdot}$};
        \def\radius{1cm}
        \draw[thick] (0,0) circle[radius=\radius];
        \node[inner sep=0pt,outer sep=0pt,square, draw,fill=white,scale=3pt] (r1) at (45:\radius) {};
        \node[inner sep=0pt,outer sep=0pt,square, draw,fill=white,scale=3pt] (l1) at (135:\radius) {};
        \node[inner sep=0pt,outer sep=0pt,square, draw,fill=white,scale=3pt] (l2) at (225:\radius) {};
        \node[inner sep=0pt,outer sep=0pt,square, draw,fill=white,scale=3pt] (r2) at (315:\radius) {};
        \draw[dashed] (-1.7,0.7)--(l1);
        \draw[dashed] (-1.7,-0.7)--(l2);
        \draw[dashed] (1.7,0.7)--(r1);
        \draw[dashed] (1.7,-0.7)--(r2);
        \draw[color=red] (-1.4,-1.2) rectangle (1.4,1.2);
    \end{tikzpicture}\,,\qquad\begin{tikzpicture}[scale=0.5,baseline=(vert_cent.base),square/.style={regular polygon,regular polygon sides=4}]
    \node (vert_cent) at (0,0) {$\phantom{\cdot}$};
    \def\radius{0.8cm}
    \draw[thick] (0,0) circle[radius=\radius];
    \node[inner sep=0pt,outer sep=0pt,square, draw,fill=white,scale=3pt] (l) at (180:\radius) {};
    \node[inner sep=0pt,outer sep=0pt,square, draw,fill=white,scale=3pt] (r) at (0:\radius) {};
    \draw[dashed] (l) to (-1.5,0);
    \draw[densely dashed] (r) to (1.5,0);
    \draw[dashed] (1.5,0) to (2,1);
    \draw[dashed] (1.5,0) to (2,-1);
    \draw[dashed] (1.5,0) to (2.2,0);
    \node[inner sep=0pt,outer sep=0pt,circle,draw,fill=black,scale=3pt] at (1.5,0) {};
    \draw[color=red] (-1.2,-1) rectangle (1.75,1);
    \end{tikzpicture}\,,\qquad\begin{tikzpicture}[scale=0.5,baseline=(vert_cent.base),square/.style={regular polygon,regular polygon sides=4}]
    \node (vert_cent) at (0,0) {$\phantom{\cdot}$};
    \def\radius{1cm}
    \draw[dashed] (90:\radius) arc (90:270:\radius);
    \draw[thick] (90:\radius) arc (90:-90:\radius);
    \node[inner sep=0pt,outer sep=0pt,circle, draw,fill=black,scale=3pt] (l) at (180:\radius) {};
    \node[inner sep=0pt,outer sep=0pt,square, draw,fill=white,scale=3pt] (r) at (0:\radius) {};
     \node[inner sep=0pt,outer sep=0pt,square, draw,fill=white,scale=3pt] (t) at (90:\radius) {};
    \node[inner sep=0pt,outer sep=0pt,square, draw,fill=white,scale=3pt] (b) at (270:\radius) {};
    \draw[dashed] (l) to (r);
    \draw[dashed] (l) to (-2.2,0);
    \draw[thick] (t) to (1.7,1);
    \draw[thick] (b) to (1.7,-1);
    \draw[color=red] (-1.4,-1.2) rectangle (1.4,1.2);
    \end{tikzpicture}\,,\qquad\begin{tikzpicture}[scale=0.5,baseline=(vert_cent.base),square/.style={regular polygon,regular polygon sides=4}]
    \node (vert_cent) at (0,0) {$\phantom{\cdot}$};
    \def\radius{0.8cm}
    \draw[dashed] (0,0) circle[radius=\radius];
    \node[inner sep=0pt,outer sep=0pt,circle, draw,fill=black,scale=3pt] (l) at (180:\radius) {};
    \node[inner sep=0pt,outer sep=0pt,circle, draw,fill=black,scale=3pt] (r) at (0:\radius) {};
    \draw[dashed] (l) to (-1.5,0);
    \draw[dashed] (l) to (r);
    \draw[densely dashed] (r) to (1.5,0);
    \draw[thick] (1.5,0) to (2,1);
    \draw[thick] (1.5,0) to (2,-1);
    \node[inner sep=0pt,outer sep=0pt,square,draw,fill=white,scale=3pt] at (1.5,0) {};
    \draw[color=red] (-1.2,-1) rectangle (1.75,1);
    \end{tikzpicture}\,,
\end{equation}
and one immediately sees that these diagrams are all primitive, containing no proper sub-divergences. The sum in (\ref{eq:gradflowrightside}) will then contain only a single term, $\gamma'=\gamma$, and the right-hand side of the gradient flow equation will look like
\begin{equation}
    G_{\circ\star}\beta^\star\supset\nc{g}{0}{}\nc{\mathfrak{b}}{n}{\gamma}\,,
\end{equation}
where the symmetry factor will necessarily be unity. Examining the left-hand side of the gradient flow equation, we note that these four diagrams in the beta function originate from two vacuum bubbles in $A$,
\begin{equation}
    \begin{tikzpicture}[scale=0.5,baseline=(vert_cent.base),square/.style={regular polygon,regular polygon sides=4}]
        \node (vert_cent) at (0,0) {$\phantom{\cdot}$};
        \node (center) at (0,0) {};
        \def\radius{1cm}
        \draw[thick] (center) circle[radius=\radius];
        \node[inner sep=0pt,outer sep=0pt,square, draw,fill=white,scale=3pt] (t) at (90:\radius) {};
        \node[inner sep=0pt,outer sep=0pt,square, draw,fill=white,scale=3pt] (l) at (180:\radius) {};
        \node[inner sep=0pt,outer sep=0pt,square, draw,fill=white,scale=3pt] (b) at (270:\radius) {};
        \node[inner sep=0pt,outer sep=0pt,square, draw,fill=white,scale=3pt] (r) at (0:\radius) {};
        \node[inner sep=0pt,outer sep=0pt,circle, draw,fill=black,scale=3pt] (c) at (0,0) {};
        \draw[densely dashed] (t) to (c);
        \draw[densely dashed] (b) to (c);
        \draw[densely dashed] (r) to (c);
        \draw[densely dashed] (l) to (c);
    \end{tikzpicture}\,,\qquad\begin{tikzpicture}[scale=0.5,baseline=(vert_cent.base),square/.style={regular polygon,regular polygon sides=4}]
        \node (vert_cent) at (0,0) {$\phantom{\cdot}$};
        \node (center) at (0,0) {};
        \def\radius{1cm}
        \draw[dashed] (45:\radius) arc (45:315:\radius);
        \draw[dashed] (135:\radius) arc (45:-45:\radius);
        \draw[thick] (45:\radius) arc (45:-45:\radius);
        \draw[thick] (45:\radius) arc (135:225:\radius);
        \node[inner sep=0pt,outer sep=0pt,circle, draw,fill=black,scale=3pt] (l1) at (135:\radius) {};
        \node[inner sep=0pt,outer sep=0pt,circle, draw,fill=black,scale=3pt] (l2) at (225:\radius) {};
        \node[inner sep=0pt,outer sep=0pt,square, draw,fill=white,scale=3pt] (r1) at (-45:\radius) {};
        \node[inner sep=0pt,outer sep=0pt,square, draw,fill=white,scale=3pt] (r2) at (45:\radius) {};
        \draw[dashed] (l1) to (l2);
    \end{tikzpicture}\,,
\end{equation}
so that we will have to solve four equations with only $\nc{a}{3}{2}$, $\nc{a}{3}{3}$ and $\nc{g}{0}{}$, clearly necessitating a constraint. One finds in particular that the normalisation of the fermionic part of the metric must be
\begin{equation}\label{eq:g00fixing}
    \nc{g}{0}{}=\frac{12\nc{\mathfrak{b}}{2}{2}}{\nc{\mathfrak{b}}{\frac{5}{2}}{1}}=24\,,
\end{equation}
where we note that as the graphs associated with $\nc{\mathfrak{b}}{2}{2}$ and $\nc{\mathfrak{b}}{\frac{5}{2}}{1}$ are primitive, this will be a scheme-invariant statement.

With little trouble it is possible to generalise this argument to generic loop order, and construct precisely this family of constraints. Let us begin with $\overline{\Gamma}$, a primitive completion vacuum graph, by which we mean the removal of any of its vertices produces a primitive diagram, of $\text{O}(\varepsilon^n)$ with some number of $\lambda^I$ and $y^\alpha$ vertices:
\begin{equation}
    \overline{\Gamma}=\begin{tikzpicture}[baseline=(vert_cent.base),scale=0.5,square/.style={regular polygon,regular polygon sides=4}]
        \node (vert_cent) at (0,0) {$\phantom{\cdot}$};
        \draw[fill=white] (0,0) circle (1cm);
        \draw[pattern=north west lines, pattern color=black] (0,0) circle (1cm);
        \filldraw (180:1cm) circle (3pt);
        \filldraw (150:1cm) circle (3pt);
        \draw[thick, dotted] (130:1.2) arc (135:110:1.2);
        \filldraw (90:1cm) circle (3pt);
        \node[inner sep=0pt,outer sep=0pt,square, draw,fill=white,scale=3pt] at (0:1cm) {};
        \node[inner sep=0pt,outer sep=0pt,square, draw,fill=white,scale=3pt] at (-30:1cm) {};
        \node[inner sep=0pt,outer sep=0pt,square, draw,fill=white,scale=3pt] at (-90:1cm) {};
        \draw[thick, dotted] (-50:1.2) arc (-55:-80:1.2);
    \end{tikzpicture}\,.
\end{equation}
Let us suppose, for the point of illustration, that there are at least four diagrams in the beta function which can be generated by removing vertices from $\overline{\Gamma}$: two, $\Lambda_1$ and $\Lambda_2$ from the removal of a $\lambda$-vertex, and two, $\mathcal{Y}_{1}$ and $\mathcal{Y}_2$ from the removal of a $y$-vertex. If there are not four such diagrams, then $\overline{\Gamma}$ will not lead to all of the constraints outlined below. If $Q_\gamma$ gives the number of ways of of obtaining $\gamma$ by removing a vertex from $\overline{\Gamma}$, and $P_\gamma$ gives the number of distinct permutations of the free indices in $\gamma$, then the gradient flow equations associated with these four graphs will be
\begin{equation}
\begin{gathered}
    \frac{Q_{\Lambda_1}}{P_{\Lambda_1}}\nc{a}{n}{\overline{\Gamma}}=\nc{\mathfrak{b}}{n-1}{\Lambda_1}\,,\qquad\frac{Q_{\Lambda_2}}{P_{\Lambda_2}}\nc{a}{n}{\overline{\Gamma}}=\nc{\mathfrak{b}}{n-1}{\Lambda_2}\,,\\
    \frac{Q_{\mathcal{Y}_1}}{P_{\mathcal{Y}_1}}\nc{a}{n}{\overline{\Gamma}}=\nc{g}{0}{}\nc{\mathfrak{b}}{n-\frac{1}{2}}{\mathcal{Y}_1}\,,\qquad\frac{Q_{\mathcal{Y}_2}}{P_{\mathcal{Y}_2}}\nc{a}{n}{\overline{\Gamma}}=\nc{g}{0}{}\nc{\mathfrak{b}}{n-\frac{1}{2}}{\mathcal{Y}_2}\,.
\end{gathered}
\end{equation}
As $\nc{g}{0}{}$ has already been fixed by (\ref{eq:g00fixing}), these four equations will translate into three constraints. The first simply relates primitive diagrams from the same beta function,
\begin{equation}
    \frac{P_{\Lambda_1}}{Q_{\Lambda_1}}\nc{\mathfrak{b}}{n-1}{\Lambda_1}=\frac{P_{\Lambda_2}}{Q_{\Lambda_2}}\nc{\mathfrak{b}}{n-1}{\Lambda_2}\,,\qquad\frac{P_{\mathcal{Y}_1}}{Q_{\mathcal{Y}_1}}\nc{\mathfrak{b}}{n-\frac{1}{2}}{\mathcal{Y}_1}=\frac{P_{\mathcal{Y}_2}}{Q_{\mathcal{Y}_2}}\nc{\mathfrak{b}}{n-\frac{1}{2}}{\mathcal{Y}_2}\,,
\end{equation}
while the third relates a scalar primitive diagram to a fermionic primitive diagram
\begin{equation}
    \frac{P_{\Lambda_1}Q_{\mathcal{Y}_1}}{P_{\mathcal{Y}_1}Q_{\Lambda_1}}\frac{\nc{\mathfrak{b}}{n-1}{\Lambda_1}}{\nc{\mathfrak{b}}{n-\frac{1}{2}}{\mathcal{Y}_1}}=\frac{12\nc{\mathfrak{b}}{2}{2}}{\nc{\mathfrak{b}}{\frac{5}{2}}{1}}=24\,.
\end{equation}
Note that of course one can always take linear combinations of these constraints. These relations generalise the purely scalar primitive constraints identified, e.g., in \cite{Pannell:2024sia}. In the multiscalar case, it has been proven using inversion properties of the associated integrals that these constraints will be satisfied to all loop orders\cite{Wallace:1974dy,Schnetz:2008mp,Panzer:2016snt}. As the symmetry of the metric was not used, these constraints will follow from the Weyl consistency conditions \cite{Jack:1990eb, Osborn:1991gm}, and thus must be obeyed at all loop orders, even when fermions are involved. However, it is not clear if there is a similar proof directly on the level of the amplitudes for the constraints which involve removing one or more $y$-vertex. For the specific case of $\overline{\Gamma}=\begin{tikzpicture}[scale=0.5,baseline=(vert_cent.base),square/.style={regular polygon,regular polygon sides=4}]
        \node (vert_cent) at (0,0) {$\phantom{\cdot}$};
        \node (center) at (0,0) {};
        \def\radius{1cm}
        \draw[dashed] (45:\radius) arc (45:315:\radius);
        \draw[dashed] (135:\radius) arc (45:-45:\radius);
        \draw[thick] (45:\radius) arc (45:-45:\radius);
        \draw[thick] (45:\radius) arc (135:225:\radius);
        \node[inner sep=0pt,outer sep=0pt,circle, draw,fill=black,scale=3pt] (l1) at (135:\radius) {};
        \node[inner sep=0pt,outer sep=0pt,circle, draw,fill=black,scale=3pt] (l2) at (225:\radius) {};
        \node[inner sep=0pt,outer sep=0pt,square, draw,fill=white,scale=3pt] (r1) at (-45:\radius) {};
        \node[inner sep=0pt,outer sep=0pt,square, draw,fill=white,scale=3pt] (r2) at (45:\radius) {};
        \draw[dashed] (l1) to (l2);
    \end{tikzpicture}$, one can easily determine
\begin{equation}
    P_{\Lambda_1}=4\,,\qquad P_{\mathcal{Y}_1}=1\,,\qquad Q_{\Lambda_1}=2\,,\qquad Q_{\mathcal{Y}_1}=2\,,
\end{equation}
to correctly obtain the first of the constraints in (\ref{eq:12constraints}). We note that of the constraints listed in Table \ref{tab:constraints}, one of the constraints at 1/2 loop, five of the constraints at 2/3 loop and 62 of the constraints at 3/4 loop are of this primitive-primitive type. While careful analysis of the other equations may permit them to be placed in similar families which appear at other loop orders, one sees that their form is considerably more complicated, and thus their origin is murkier.

As a final comment before moving on to solving the gradient flow equation at higher loop orders, we note that the solution to (\ref{eq:eqns12}) requires that the metric $G_{\alpha\beta}$ receive non-trivial corrections even at next-to leading order. This is in contrast to the multiscalar case, where the corrections to the metric were only necessary beginning at three-loops, or next-to-next-to leading order. This result is still consistent with the existence of Riemann normal coordinates, because the lowest order term is $\text{O}(y^2)$ not $\text{O}(y)$.

At higher orders the principle of generating and solving the equations remains the same, though as one can see from Table \ref{tab:graphnum} the number of equations will grow rapidly. We find that satisfying the gradient flow equations through 3/4 loop order in $d=4$ is equivalent to 955 constraint equations on the beta function, as summarised in Table \ref{tab:constraints}.
\begin{table}[H]
    \centering
    \begin{tabular}{|c|c|c|c|c|}
        \hline
        Loop order in $\beta_\lambda$/$\beta_y$: & 0/1 & 1/2 & 2/3 & 3/4  \\
        \hline
        Constraints on $\beta$ & -- & 4+(0) & 63+(4) & 888+(57)\\
        Constraints on $S$ and $P$ & -- & -- & 4+(0) & 63+(2) \\
        \hline
    \end{tabular}
    \caption{The number of constraints on the coefficients in the beta function and in the beta shift appearing at a specified loop order, in $d=4$ and $d=4-\varepsilon$. The listed numbers $n+(m)$ indicate that there will be $n$ constraints in $d=4$ and an additional $m$ constraints in $d=4-\varepsilon$. Note that because of the loop order mixing, the loop order is listed in the form $n/n+1$, where $n$ is the loop order in the scalar beta function.}
    \label{tab:constraints}
\end{table}
Beginning at 2/3 loop order, the $\beta$ function and the $B$ function will differ, and we must ask if either is compatible with gradient flow. There are five three loop diagrams contributing to the beta shift, two diagrams in $S_{ij}$,
\begin{equation}
    (\nc{S}{3}{})_{ij}=\nc{S}{3}{1}\,\begin{tikzpicture}[scale=0.5,baseline=(vert_cent.base),square/.style={regular polygon,regular polygon sides=4}]
        \node (vert_cent) at (0,0) {$\phantom{\cdot}$};
        \def\radius{1cm}
        \draw[thick] (120:\radius) arc (120:-90:\radius);
        \draw[dashed] (120:\radius) arc (120:270:\radius);
        \node[inner sep=0pt,outer sep=0pt,circle, draw,fill=black,scale=3pt] (l) at (180:\radius) {};
        \node[inner sep=0pt,outer sep=0pt,square, draw,fill=white,scale=3pt] (r) at (0:\radius) {};
        \node[inner sep=0pt,outer sep=0pt,square, draw,fill=white,scale=3pt] (b) at (270:\radius) {};
        \node[inner sep=0pt,outer sep=0pt,square, draw,fill=white,scale=3pt] (tl) at (120:\radius) {};
        \node[inner sep=0pt,outer sep=0pt,square, draw,fill=white,scale=3pt] (tr) at (60:\radius) {};
        \draw[thick] (tl) to (b);
        \draw[dashed] (tr) to (l);
        \draw[dashed] (-1.7,0)--(l);
        \draw[dashed] (1.7,0)--(r);
        \node[yshift=7pt,xshift=-3pt] at (-1.7,0) {$i$};
        \node[yshift=7pt,xshift=3pt] at (1.7,0) {$j$};
    \end{tikzpicture}+\nc{S}{3}{2}\,\begin{tikzpicture}[scale=0.5,baseline=(vert_cent.base),square/.style={regular polygon,regular polygon sides=4}]
        \node (vert_cent) at (0,0) {$\phantom{\cdot}$};
        \def\radius{1cm}
        \draw[thick] (0,0) circle[radius=\radius];
        \draw[densely dashed] (150:\radius) arc (-90:-30:\radius);
        \node[inner sep=0pt,outer sep=0pt,square, draw,fill=white,scale=3pt] (l) at (180:\radius) {};
        \node[inner sep=0pt,outer sep=0pt,square, draw,fill=white,scale=3pt] (r) at (0:\radius) {};
        \node[inner sep=0pt,outer sep=0pt,square, draw,fill=white,scale=3pt] (b) at (270:\radius) {};
        \node[inner sep=0pt,outer sep=0pt,square, draw,fill=white,scale=3pt] (t) at (90:\radius) {};
        \node[inner sep=0pt,outer sep=0pt,square, draw,fill=white,scale=3pt] (tl) at (150:\radius) {};
        \node[inner sep=0pt,outer sep=0pt,square, draw,fill=white,scale=3pt] (tr) at (45:\radius) {};
        \draw[dashed] (tr) to (b);
        \draw[dashed] (-1.7,0)--(l);
        \draw[dashed] (1.7,0)--(r);
        \node[yshift=7pt,xshift=-3pt] at (-1.7,0) {$i$};
        \node[yshift=7pt,xshift=3pt] at (1.7,0) {$j$};
    \end{tikzpicture}\,,
\end{equation}
and then three diagrams in $P_{ab}$,
\begin{equation}
    (\nc{P}{3}{})_{ab}=\nc{P}{3}{1}\,\begin{tikzpicture}[scale=0.5,baseline=(vert_cent.base),square/.style={regular polygon,regular polygon sides=4}]
        \node (vert_cent) at (0,0) {$\phantom{\cdot}$};
        \def\radius{1cm}
        \draw[thick] (180:\radius) arc (180:270:\radius);
        \draw[densely dashed] (270:\radius) arc (270:360:\radius);
        \draw[thick] (0:\radius) arc (0:45:\radius);
        \draw[densely dashed] (45:\radius) arc (45:90:\radius);
        \draw[thick] (90:\radius) arc (90:150:\radius);
        \draw[densely dashed] (150:\radius) arc (150:180:\radius);
        \draw[thick] (150:\radius) arc (-90:-30:\radius);
        \node[inner sep=0pt,outer sep=0pt,square, draw,fill=white,scale=3pt] (l) at (180:\radius) {};
        \node[inner sep=0pt,outer sep=0pt,square, draw,fill=white,scale=3pt] (r) at (0:\radius) {};
        \node[inner sep=0pt,outer sep=0pt,square, draw,fill=white,scale=3pt] (b) at (270:\radius) {};
        \node[inner sep=0pt,outer sep=0pt,square, draw,fill=white,scale=3pt] (t) at (90:\radius) {};
        \node[inner sep=0pt,outer sep=0pt,square, draw,fill=white,scale=3pt] (tl) at (150:\radius) {};
        \node[inner sep=0pt,outer sep=0pt,square, draw,fill=white,scale=3pt] (tr) at (45:\radius) {};
        \draw[thick] (tr) to (b);
        \draw[thick] (-1.7,0)--(l);
        \draw[thick] (1.7,0)--(r);
        \node[yshift=7pt,xshift=-3pt] at (-1.7,0) {$a$};
        \node[yshift=7pt,xshift=3pt] at (1.7,0) {$b$};
    \end{tikzpicture}+\nc{P}{3}{2}\,\begin{tikzpicture}[scale=0.5,baseline=(vert_cent.base),square/.style={regular polygon,regular polygon sides=4}]
        \node (vert_cent) at (0,0) {$\phantom{\cdot}$};
        \def\radius{1cm}
        \draw[densely dashed] (0:\radius) arc (0:45:\radius);
        \draw[thick] (45:\radius) arc (45:90:\radius);
        \draw[densely dashed] (90:\radius) arc (90:135:\radius);
        \draw[thick] (135:\radius) arc (135:180:\radius);
        \draw[densely dashed] (180:\radius) arc (180:270:\radius);
        \draw[thick] (270:\radius) arc (270:360:\radius);
        \node[inner sep=0pt,outer sep=0pt,square, draw,fill=white,scale=3pt] (l) at (180:\radius) {};
        \node[inner sep=0pt,outer sep=0pt,square, draw,fill=white,scale=3pt] (r) at (0:\radius) {};
        \node[inner sep=0pt,outer sep=0pt,square, draw,fill=white,scale=3pt] (b) at (270:\radius) {};
        \node[inner sep=0pt,outer sep=0pt,square, draw,fill=white,scale=3pt] (t) at (90:\radius) {};
        \node[inner sep=0pt,outer sep=0pt,square, draw,fill=white,scale=3pt] (tl) at (135:\radius) {};
        \node[inner sep=0pt,outer sep=0pt,square, draw,fill=white,scale=3pt] (tr) at (45:\radius) {};
        \draw[thick] (t) to (b);
        \draw[thick] (tr) to (tl);
        \draw[thick] (-1.7,0)--(l);
        \draw[thick] (1.7,0)--(r);
        \node[yshift=7pt,xshift=-3pt] at (-1.7,0) {$a$};
        \node[yshift=7pt,xshift=3pt] at (1.7,0) {$b$};
    \end{tikzpicture}+\nc{P}{3}{3}\,\begin{tikzpicture}[scale=0.5,baseline=(vert_cent.base),square/.style={regular polygon,regular polygon sides=4}]
        \node (vert_cent) at (0,0) {$\phantom{\cdot}$};
        \def\radius{1cm}
        \draw[densely dashed] (180:\radius) arc (180:270:\radius);
        \draw[thick] (270:\radius) arc (270:360:\radius);
        \draw[densely dashed] (0:\radius) arc (0:45:\radius);
        \draw[thick] (45:\radius) arc (45:180:\radius);
        \draw[densely dashed] (150:\radius) arc (-90:-30:\radius);
        \node[inner sep=0pt,outer sep=0pt,square, draw,fill=white,scale=3pt] (l) at (180:\radius) {};
        \node[inner sep=0pt,outer sep=0pt,square, draw,fill=white,scale=3pt] (r) at (0:\radius) {};
        \node[inner sep=0pt,outer sep=0pt,square, draw,fill=white,scale=3pt] (b) at (270:\radius) {};
        \node[inner sep=0pt,outer sep=0pt,square, draw,fill=white,scale=3pt] (t) at (90:\radius) {};
        \node[inner sep=0pt,outer sep=0pt,square, draw,fill=white,scale=3pt] (tl) at (150:\radius) {};
        \node[inner sep=0pt,outer sep=0pt,square, draw,fill=white,scale=3pt] (tr) at (45:\radius) {};
        \draw[thick] (tr) to (b);
        \draw[thick] (-1.7,0)--(l);
        \draw[thick] (1.7,0)--(r);
        \node[yshift=7pt,xshift=-3pt] at (-1.7,0) {$a$};
        \node[yshift=7pt,xshift=3pt] at (1.7,0) {$b$};
    \end{tikzpicture}\,.
\end{equation}
Plugging in this shift, and comparing the constraints for $\beta$ and $B$, one finds that the $\overline{\text{MS}}$ values of the beta function coefficients do not satisfy 11 of the 63 constraints appearing at 2/3 loop order, while for $B$ these 11 equations become four constraints on $S_{ij}$ and $P_{ab}$:
\begin{equation}
    \nc{S}{3}{1}=\tfrac{5}{8}\,,\qquad\nc{S}{3}{2}=\tfrac{3}{4}\,,\qquad2\nc{P}{3}{1}+\nc{P}{3}{2}-2\nc{P}{3}{3}=\tfrac{5}{8}\,,\qquad \nc{P}{3}{2}-4\nc{P}{3}{3}=-\tfrac{7}{8}\,.
\end{equation}
An explicit calculation of this beta shift has been performed at three loops\cite{Fortin:2012hn,Davies:2021mnc,Jack:2024sjr}, with the coefficients being
\begin{equation}\label{eq:Slist}
    \nc{S}{3}{1}=\tfrac{5}{8}\,,\qquad \nc{S}{3}{2}=\tfrac{3}{4}\,,\qquad\nc{P}{3}{1}=\tfrac{7}{16}\,,\qquad\nc{P}{3}{2}=\tfrac{3}{8}\,,\qquad\nc{P}{3}{3}=\tfrac{5}{16}\,.
\end{equation}
Note that the factor of 2 in $\nc{S}{3}{2}$ relative to that listed in (\ref{eq:Sthreeloops}) arises from the transition from complex couplings to real couplings. Using these values one finds that the three remaining constraints are all satisfied, and thus that the $B$ function remains gradient even when $\beta$ is not. One sees that the inclusion of the beta shift is essential to properly constructing RG monotonicity theorems. We note that the failure of gradiency has occurred here as soon as the shift could have an effect, unlike in the case of purely scalar theories where the shift begins at five loops but did not appear in the constraints due to cancellations until six loops.

At four loops, there are an additional 38 diagrams appearing in $S_{ij}$ and 61 digrams appearing in $P_{ab}$, exhibited in Appendix \ref{app:SPcon}. Plugging in the shift and the $\overline{\text{MS}}$ values for the $\nc{b}{n}{m}$, one finds that 472 of 888 the constraints are not satisfied for $\beta$, turning into 63 constraints on the beta shift, as captured in Table \ref{tab:constraints}. Of these 63 constraints, which are listed in Appendix \ref{app:SPcon}, 20 are noteworthy for fixing individual coefficients exactly,
\begin{equation}
\begin{gathered}
    \nc{P}{4}{19}=0\,,\qquad\nc{P}{4}{20}=-\tfrac{3}{8}\,,\qquad\nc{P}{4}{45}=0\,,\qquad\nc{P}{4}{47}=0\,,\qquad\nc{S}{4}{17}=0\,,\\\nc{P}{4}{4}=0\,,\qquad\nc{P}{4}{7}=\tfrac{\pi ^4}{120}-\tfrac{1}{3}\,,\qquad\nc{S}{4}{2}=\tfrac{7}{24}+\tfrac{\pi
   ^4}{120}\,,\qquad\nc{P}{4}{52}=\tfrac{3}{8}\,,\qquad\nc{P}{4}{53}=0\,,\\\nc{P}{4}{54}=0\,,\qquad\nc{S}{4}{31}=\tfrac{5}{4}\,,\qquad\nc{P}{4}{58}=-\tfrac{1}{3}\,,\qquad\nc{S}{4}{34}=\tfrac{55}{48}\,,\qquad\nc{P}{4}{56}=\tfrac{43}{96}\,,\qquad\nc{S}{4}{27}=0\,,\\\nc{P}{4}{61}=\tfrac{1}{6}\,,\qquad\nc{S}{4}{37}=\tfrac{13}{16}\,,\qquad\nc{S}{4}{36}=\tfrac{23}{32}\,,\qquad\nc{S}{4}{38}=\tfrac{13}{96}\,.
\end{gathered}
\end{equation}
To verify that these are indeed satisfied, one would need to perform an explicit calculation of $S$ and $P$ at four loops. Given the immense amount of cancellation that has already had to occur in the gradient flow equation to reduce to these constraints, it seems as though it would be an immense conspiracy were such a calculation to not agree with these constraints, and thus that the weight of evidence points to the $B$-functions being genuinely gradient through 3/4 loop order.

\subsection{Extension to \texorpdfstring{$d=4-\varepsilon$}{d=4-epsilon}}
The fixed points which will correspond to three-dimensional CFTs do not exist in exactly four dimensions, and consequently we are really only interested in perturbative RG flows for dimensions slightly less than four. We will thus seek to expand the $d=4$ solution outlined in the last section to a solution in $d=4-\varepsilon$, following the same procedure used in purely scalar theories\cite{Pannell:2024sia}. The structure of the equations will remain the same, all that needs to be modified with respect to the previous section is the inclusion of explicit $\varepsilon$-dependence in $\beta^\star$. In $\overline{\text{MS}}$, this $\varepsilon$-dependence appears only via the inclusion of terms linear in the couplings coming from the classical dimension of the perturbing operators,
\begin{equation}
    \beta^I\supset-\varepsilon\lambda^I\,,\qquad\beta^\alpha\supset-\tfrac12\varepsilon y^\alpha\,.
\end{equation}
However, in a generic scheme there may also be $\varepsilon$-dependence in the other coefficients of the beta function, and thus to be fully general we will write
\begin{equation}
    \beta^\star=\sum_n\sum_m(\nc{\mathfrak{b}}{n}{m}+\varepsilon\nc{\mathfrak{e}}{n}{m})(\nc{\beta}{n}{m})^\star\,.
\end{equation}
To compensate for this new $\varepsilon$-dependence, we must introduce factors of $\varepsilon$ into $A$ and $G_{\circ\star}$ as well. Given that we have a solution which is valid when $\varepsilon\rightarrow0$, we wish to introduce these factors such that this limit is respected. Consequently, we write
\begin{equation}
\begin{split}
    A&=\sum_n\sum_m(\nc{a}{n}{m}+\varepsilon\lsp\nc{d}{n}{m}(\varepsilon))\nc{A}{n}{m}\,, \\
    G_{\circ\star}&=\sum_n\sum_m(\nc{g}{n}{m}+\varepsilon\lsp\nc{h}{n}{m}(\varepsilon))(\nc{G}{n}{m})_{\circ\star}\,.
\end{split}
\end{equation}
Here, $\nc{a}{n}{m}$ and $\nc{g}{n}{m}$ are numbers independent of the dimension, given by the four-dimensional solution to gradient flow, while $\nc{d}{n}{m}(\varepsilon)$ and $\nc{h}{n}{m}(\varepsilon)$ are polynomials in $\varepsilon$ with arbitrary coefficients we must solve for.

Schematically, the gradient flow equations then become
\begin{equation}
\begin{gathered}
    \varepsilon(-\lambda_I+H_{I\star}(\varepsilon)\beta^\star+G_{I\star}\mathfrak{E}^*(\varepsilon))+\varepsilon^2H_{I\star}(\varepsilon)\mathfrak{E}^*(\varepsilon)=\varepsilon\partial_I D(\varepsilon)\,, \\
    \varepsilon(-\tfrac{1}{2}y_\alpha+H_{\alpha\star}(\varepsilon)\beta^\star+G_{\alpha\star}\mathfrak{E}^*(\varepsilon))+\varepsilon^2H_{\alpha\star}(\varepsilon)\mathfrak{E}^*(\varepsilon)=\varepsilon\partial_\alpha D(\varepsilon)\,,
\end{gathered}
\end{equation}
where we have used the solution in $d=4$ to remove the $\text{O}(\varepsilon^0)$ terms, and where $\mathfrak{E}^*$ includes the $\nc{e}{n}{m}$. Analysis of these equations then proceeds exactly as in the previous section, though we must be careful to ensure that no poles in $\varepsilon$ appear in $D(\varepsilon)$ and $H_{\circ\star}(\varepsilon)$, which would spoil the $d\rightarrow 4$ limit. The linear terms in the beta function can quite easily be absorbed into the $A$-function via
\begin{equation}
    A\supset-\frac{1}{2}\varepsilon\,\begin{tikzpicture}[scale=0.5,baseline=(vert_cent.base)]
        \node (vert_cent) at (0,0) {$\phantom{\cdot}$};
        \draw[dashed] (0,0) circle (1cm);
        \draw[dashed] (1,0) to[out=135, in=45] (-1,0);
        \draw[dashed] (1,0) to[out=225, in=315] (-1,0);
        \node[inner sep=0pt,outer sep=0pt,circle, draw,fill=black,scale=3pt] at (-1,0) {};
        \node[inner sep=0pt,outer sep=0pt,circle, draw,fill=black,scale=3pt] at (1,0) {};
    \end{tikzpicture}-\frac{1}{4}\varepsilon\nc{g}{0}{}\,\begin{tikzpicture}[scale=0.5,baseline=(vert_cent.base),square/.style={regular polygon,regular polygon sides=4}]
        \node (vert_cent) at (0,0) {$\phantom{\cdot}$};
        \draw[dashed] (-1,0) arc (180:0:1cm);
        \draw[thick] (-1,0) arc (180:360:1cm);
        \draw[thick] (-1,0) -- (1,0);
        \node[inner sep=0pt,outer sep=0pt,square, draw,fill=white,scale=3pt] at (-1,0) {};
        \node[inner sep=0pt,outer sep=0pt,square, draw,fill=white,scale=3pt] at (1,0) {};
    \end{tikzpicture}\,,
\end{equation}
or in the above language, taking $\nc{d}{2}{1}=-\tfrac{1}{2}\varepsilon$ and $\nc{d}{1}{}=-\tfrac{1}{4}\varepsilon\nc{g}{0}{}$. At $n$ loops, the linear terms in the beta function will pull down metric diagrams which would have first appeared at $n+1$ loops in $d=4$. As some of these will have been fixed in terms of the beta function coefficients at $n+1$ loops, constraints from the $\varepsilon$-part of the $n$ loop gradient flow equation will produce constraint equations at $n+1$ loops. Thus, as we are interested in gradient flow through 3/4 loops, we only need to consider the $\varepsilon$-part of the gradient flow equation through 2/3 loops. The $\varepsilon$-part at 3/4 loops will include metric coefficients $\nc{g}{4}{m}$ which are unfixed, and can thus be used to solve the resulting equations without the need for additional constraints. While this may conflict with a putative $d=4$ solution at 4/5 loops, and then produce additional constraints, it is sufficient for our purposes.

We find that a consistent solution for the $\nc{d}{n}{m}(\varepsilon)$ and $\nc{h}{n}{m}(\varepsilon)$ can be found through 3/4 loops, but that the beta function must then obey an additional 61 constraints, as outlined in Table \ref{tab:constraints}. While obeying the 4 constraints at 2/3 loop order, we find that the $\overline{\text{MS}}$ beta function fails to obey 5 of the 57 constraints at 3/4 loops. As in $d=4$, it is possible for the beta shifts to rectify this situation, as long as $S_{ij}$ satisfies the two new constraints
\begin{equation}
    \nc{S}{4}{28}-\tfrac14\nc{S}{4}{29}=-\tfrac{13}{24}\,,\qquad\tfrac14\nc{S}{4}{4}-\nc{S}{4}{6}-\tfrac{1}{16}\nc{S}{4}{8}+\tfrac14\nc{S}{4}{11}=-\tfrac{1}{12}\,.
\end{equation}
While we expect that these equations are obeyed, a true check would require the calculation of these coefficients at four loops.

\subsection{Scheme invariance of the constraints}
The physics of a QFT cannot depend upon the particular renormalisation scheme which is chosen for the purposes of calculation. As we want to ascribe physical reality to gradient flow, even though $A$, $G_{\circ\star}$ and $\beta^\star$ will not be directly physical for generic $\lambda^I$ and $y^\alpha$, the constraints on the beta function coefficients must be scheme invariant. That being the case, the fact that they are satisfied by $B^\star$ in $\overline{\text{MS}}$ ensures that they will be satisfied in any scheme. A generic scheme change can be realised as a diffeomorphism on the couplings $\{\lambda^I,y^\alpha\}$, which we will write as
\begin{equation}\label{eq:diff}
    \lambda^I\rightarrow \lambda'^{\,I}=\lambda^I+\sum_{n=2}^{\infty}\sum_m \nc{r}{n}{m}\nc{T}{n}{m}^I(\lambda,y)\,,\qquad y^\alpha\rightarrow y'^{\,\alpha}=y^\alpha+\sum_{n=\frac{3}{2}}^\infty\sum_m\nc{r}{n}{m}\nc{\mathfrak{T}}{n}{m}^\alpha(\lambda,y)\,.
\end{equation}
As we have already summed over all possible diagrams in constructing the beta function, the change of scheme will not introduce additional diagrams but only change the coefficients of those diagrams, so that
\begin{equation}
    \beta'^{\llsp\star}=\sum_{n}\sum_m\nc{{\mathfrak{b}'}}{n}{\!\!m}(\nc{\beta}{n}{m})^*(\lambda',y')\,,
\end{equation}
where now the vertices inside the diagrams are primed. These beta functions can alternatively be determined from (\ref{eq:diff}) by acting with a derivative with respect to $\ln\mu$,
\begin{equation}\label{eq:schemechangebeta}
\begin{split}
    \beta'^{\llsp I}(\lambda',y')=\frac{d\lambda'^{\lsp I}}{d\ln\mu}=\beta^I(\lambda,y)+\sum_{n=2}^\infty\sum_m\sum_{\hat{T}\subset T}\nc{r}{n}{m}(\nc{\hat{T}}{n}{m})^{I}_{\,J}\beta^J(\lambda,y)\,,\\
    \beta'^{\llsp\alpha}(\lambda',y')=\frac{dy'^{\llsp\alpha}}{d\ln\mu}=\beta^\alpha(\lambda,y)+\sum_{n=\frac{3}{2}}^\infty\sum_m\sum_{\hat{\mathfrak{T}}\subset \mathfrak{T}}\nc{r}{n}{m}(\nc{\hat{\mathfrak{T}}}{n}{m})^{\alpha}_{\,\beta}\beta^\beta(\lambda,y)\,.
\end{split}
\end{equation}
Equating these two representations of the beta functions, and expanding the primed coordinates using (\ref{eq:diff}) produces a set of linear equations determining the $\nc{{\mathfrak{b}'}}{n}{\!\!m}$ in terms of $\{\nc{\mathfrak{b}}{n}{m},\nc{r}{n}{m}\}$ order by order in $n$. As our derivation of the constraint equations was fully general, to test their scheme invariance all we must do is plug in the scheme changed $\nc{{\mathfrak{b}'}}{n}{\!\!m}$, and track the scheme change parameters $\nc{r}{n}{m}$. Doing so, one finds that the scheme change parameters completely drop out in all of the constraints we have found, so that they are correctly scheme invariant. Thus, their satisfaction in $\overline{\text{MS}}$ implies that they are satisfied in all schemes, and that gradient flow holds physically.

One proceeds in precisely the same way for the $\nc{e}{n}{m}$, but now including the linear terms proportional to $\varepsilon$. The $\varepsilon$-dependent terms in (\ref{eq:schemechangebeta}) will be
\begin{equation}
\begin{split}
    \beta'^{\llsp I}\supset-\varepsilon\lambda^I-\sum_{n=2}^{\infty}\sum_mn\nc{r}{n}{m}\nc{T}{n}{m}^I
(\lambda,y)\,,\\
\beta'^{\llsp\alpha}\supset-\tfrac{1}{2}\varepsilon y^\alpha-\sum_{n=\frac{3}{2}}^{\infty}\sum_mn\nc{r}{n}{m}\nc{T}{n}{m}^I
(\lambda,y)\,,\end{split}
\end{equation}
which we must now match with
\begin{equation}
    \beta'^{\llsp\star}\supset-\Delta\varepsilon g'^{\llsp*}+\sum_{n}\sum_m\nc{{\mathfrak{e}'}}{n}{m}(\nc{\beta}{n}{m})^*(\lambda',y')\,,
\end{equation}
where $-\Delta\varepsilon g'^{\llsp\alpha}=-\frac{1}{2}\varepsilon y'^{\llsp\alpha}$ and $-\Delta\varepsilon g'^{\llsp I}=-\varepsilon\lambda'^{\llsp I}$. Solving for the $\nc{{\mathfrak{e}'}}{n}{m}$, one finds that scheme invariance extends to the additional constraints found in $d=4-\varepsilon$.

\subsection{Matching \texorpdfstring{$A$}{A} with \texorpdfstring{$\widetilde{F}$}{F-tilde}}
Although finding the set of constraint equations from (\ref{eq:GradFloweqn}) requires fixing as many of the $\nc{a}{n}{m}$ and $\nc{g}{n}{m}$ as possible, after the dust settles there still exists a solution space for $A$ and the metric. Up to this point, these objects have been mathematical abstractions with no meaningful physical interpretation, and we would thus like to use the remaining freedom to match $A$ with an observable quantity which is computable in the field theory. In \cite{Giombi:2014xxa} it was conjectured that there exists a quantity derived from the free energy of the theory placed on the $d$-sphere,
\begin{equation}
    \widetilde{F}=-\sin{\left(\frac{d\lsp\pi}{2}\right)}F\,,
\end{equation}
which is weakly monotonic for RG flows in all $d$, and moreover becomes manifestly equal to $c$, $F$ and $A$ in two, three and four dimensions respectively. This $\widetilde{F}$-theorem has been verified perturbatively to low loop orders in $d=4-\varepsilon$ for multiscalar theories\cite{Fei:2015oha} as well as scalar-fermion theories\cite{Fei:2016sgs}, and in \cite{Pannell:2025ixz} two of the authors found that for purely scalar theories $A$ could be fixed such that
\begin{equation}\label{eq:FAmatchscalar}
    \widetilde{F}|_{\text{CFT}}-N_s\widetilde{F}_{s}=\frac{\pi}{288}A(\lambda^*_{\text{CFT}})
\end{equation}
at any fixed point, where $\widetilde{F}_s$ is a constant shift associated with the free energy of the free scalar. As $\widetilde{F}$ is defined only at the disparate CFT points in theory space, matching to $A$ provides a functional extension away from the fixed points upgrading the weak conjecture to a conjecture of full gradient flow.

It is then natural to ask whether or not the introduction of Yukawa couplings breaks this relationship between the $A$ we compute and $\widetilde{F}$. Importantly, all purely scalar fixed points can be realised as fixed points of the mixed scalar-fermion system, with $y_{iab}=0$. The addition of $N$ free fermionic fields will only shift $\widetilde{F}$ by the constant $N\widetilde{F}_f$, so that we will want to demand that (\ref{eq:FAmatchscalar}) carries through with exactly the same proportionality of $\pi/288$, even at fixed points with $y^*_{iab}\neq0$.

The value of $\widetilde{F}$ for the Gross--Neveu--Yukawa CFT was computed in \cite{Fei:2016sgs} to be
\begin{equation}\label{eq:FGNY}
\begin{split}
    \widetilde{F}_{\text{GNY}}=\widetilde{F}_s+N\widetilde{F}_f-\frac{N\pi}{96(N+6)}\varepsilon^2-\frac{1}{31104(N+6)^3}\bigg(161N^3+3690N^2+11880N+216\\+(N^2+132N+36)\sqrt{N^2+132N+36}\bigg)\pi\varepsilon^3+\text{O}(\varepsilon^4)\,.
\end{split}
\end{equation}
To see if it is possible to choose the coefficients left unfixed by the solution to gradient flow such that this matches the value of $A$ at that point, we must plug in the critical values for $\lambda^I$ and $y^\alpha$. As we are matching with $\widetilde{F}$ through $\text{O}(\varepsilon^3)$, we will only need the leading order values of these coefficients.\footnote{One might think that we would also need the next-to-leading order contribution to $y^\alpha$, but because the 0/1 loop part of $A$ is exactly the integral of the one loop $\beta^\alpha$, these terms will be proportional the one-loop beta function, which vanishes at fixed points.} There is only a single scalar field in this model, so that
\begin{equation}
    \lambda^I\to\lambda^{(1)}\varepsilon+\text{O}(\varepsilon^2)\,.
\end{equation}
The Yukawa tensor can be written as
\begin{equation}
    y^\alpha\to\big(y^{(1)}\varepsilon^{\frac{1}{2}}+\text{O}(\varepsilon^{\frac{3}{2}})\big)d_{ab}\,,
\end{equation}
where $d_{ab}$ is the matrix given in \cite[Equation (3.3)]{Pannell:2023tzc} without the factors of $y$. The only essential feature of this matrix we will use to compute $A$ is that it squares to the identity,
\begin{equation}
    d_{ab}d_{bc}=I_{ac}\,.
\end{equation}
The critical values of the couplings are then\cite{Fei:2016sgs,Pannell:2023tzc}
\begin{equation}\label{eq:GNYcrit}
    y^{(1)}=\frac{1}{\sqrt{N+6}}\,,\qquad\qquad \lambda^{(1)}=\frac{6-N+\sqrt{N^2+132N+36}}{6(N+6)}\,,
\end{equation}
where $N=2N_{\text{Weyl}}=4N_{\text{Dirac}}$ gives the number of 3d Majorana spinors in the CFT. Plugging this into (\ref{eq:Adef}) and collecting all terms up through $\text{O}(\varepsilon^3)$ one finds
\begin{equation}
\begin{split}
    \frac{\pi}{288}\lsp A_{\text{GNY}}=&-\frac{N\pi}{96(N+6)}\varepsilon^2-\frac{1}{31104(N+6)^3} \bigg((N^2+132N+36)\sqrt{N^2+132N+36}\\&+2 N (-N (95 N+261)-378)+216+54 N (N+6)^2 (\nc{g}{1}{9}+6 \nc{h}{0}{2})\bigg)\pi\varepsilon^3+\text{O}(\varepsilon^4)\,.
\end{split}
\end{equation}
Where the unknown coefficients in this expression correspond to metric diagrams
\begin{equation}
    G_{IJ}\supset \varepsilon \nc{h}{0}{2}\lsp\begin{tikzpicture}[scale=0.65,
    vertex/.style={draw,circle,fill=black,minimum size=2pt,inner sep=0pt,outer sep=0pt},
    arc/.style={thick},baseline=(vert_cent.base)]
    \node (vert_cent) at (0,-0.75) {$\phantom{\cdot}$};
    \foreach [count=\i] \coord in {
(1,0), (-1,0),(1,-0.5),(-1,-0.5)}{
        \node[inner sep=0pt,outer sep=0pt] (p\i) at \coord {};
    }
    \foreach [count=\i] \coord in {
(1,-1), (-1,-1),(1,-1.5),(-1,-1.5)}{
        \node[inner sep=0pt,outer sep=0pt] (d\i) at \coord {};
    }
    \draw[dashed] (d1) edge (d2);
    \draw[dashed] (d3) edge (d4);
    \draw[dashed] (p1) edge (p2);
    \draw[dashed] (p3) edge (p4);
\end{tikzpicture}\,+\nc{g}{1}{9}\lsp\begin{tikzpicture}[scale=0.65,
    vertex/.style={draw,circle,fill=black,minimum size=3pt,inner sep=0pt},
    arc/.style={thick},baseline=(vert_cent.base)]
    \node (vert_cent) at (0,-0.75) {$\phantom{\cdot}$};
    \node[vertex] (c) at (0,-0.25) {};
    \foreach [count=\i] \coord in {
(1.00,0), (-1,0),(-1,-0.5),(1,-0.5)}{
        \node[] (p\i) at \coord {};
    }
    \foreach [count=\i] \coord in {
(-1,-1), (1,-1),(-1,-1.5),(1,-1.5)}{
        \node[] (d\i) at \coord {};
    }
    \draw[densely dashed] (c) edge (p1)
                   edge (p2)
                   edge (p3)
                   edge (p4);
    \draw[densely dashed] (d1) edge (d2);
    \draw[densely dashed] (d3) edge (d4);
\end{tikzpicture}\,.
\end{equation}
Matching this to (\ref{eq:FGNY}) requires the choice
\begin{equation}\label{eq:amatchcond}
    \nc{h}{0}{2}=\frac{1}{12}(13-2\nc{g}{1}{9})\,,
\end{equation}
which is crucially independent of the number of fields\footnote{Note that this is precisely the same condition found in \cite{Pannell:2025ixz} to match $A$ with $\widetilde{F}$ at the purely scalar $O(N)$ fixed point.}. Note that at the critical points $A$ is a scheme invariant, as it should be to match to a physical quantity like the free energy. The condition (\ref{eq:amatchcond}), however, will be correctly scheme co-variant according to the scheme transformation of $\nc{h}{0}{2}$, with the scheme variance of the right-hand side coming from $\nc{g}{1}{9}$ and beta function coefficients which we have set to their $\overline{\text{MS}}$ values.

If it is indeed possible to upgrade the $\widetilde{F}$-conjecture to a conjecture about gradient flow, this single choice should suffice to match the value of $A$ with $\widetilde{F}$ at any fixed point. At a more general fixed point, both $A$ and $\widetilde{F}$ will be series in the couplings constructed from the same tensor structures. Conditions like (\ref{eq:amatchcond}) will fix the coefficients of the tensor structures in $A$ such that they match the various integrals which appear as the coefficients in $\widetilde{F}$. Moving to a different fixed point will simply affect the value of the tensor structures without altering the coefficients, and thus change $A$ and $\widetilde{F}$ in the same way and preserve the equality between them.

For instance, at the Nambu--Jona-Lasinio--Yukawa CFT the value of $\widetilde{F}$ was also computed in \cite{Fei:2016sgs} to be
\begin{equation}
\begin{split}
    \widetilde{F}_{\text{NJLY}}=2\widetilde{F}_s+N\widetilde{F}_f-\frac{N\pi}{48(N+4)}\varepsilon^2-\frac{1}{14400(N+4)^3}\bigg(149N^3+2372N^2+1312N+64\\+(N^2+152N+16)\sqrt{N^2+152N+16}\bigg)\pi\varepsilon^3+\text{O}(\varepsilon^4)\,.
\end{split}
\end{equation}
This model now has two scalar fields, with $\lambda^I$ being given by an $O(2)$ invariant interaction
\begin{equation}
    \lambda^I\to\big(\lambda^{(1)}\varepsilon+\text{O}(\varepsilon^2)\big)(\delta_{ij}\delta_{kl}+\delta_{ik}\delta_{jl}+\delta_{il}\delta_{jk})\,.
\end{equation}
The two Yukawa tensors are then given by
\begin{equation}
    y_{1ab}\to\big(y^{(1)}\varepsilon^{\frac{1}{2}}+\text{O}(\varepsilon^{\frac{3}{2}})\big)(d_1)_{ab}\,,\qquad\qquad y_{2ab}\to\big(y^{(1)}\varepsilon^{\frac{1}{2}}+\text{O}(\varepsilon^{\frac{3}{2}})\big)(d_2)_{ab}\,,
\end{equation}
where the two matrices $d_1$ and $d_2$ are given by \cite[Equations (3.11) and (3.13)]{Pannell:2023tzc} respectively, again without the factors of $y$. The critical couplings are then given by\cite{Fei:2016sgs,Pannell:2023tzc}
\begin{equation}\label{eq:NJLYcrit}
    y^{(1)}=\frac{1}{\sqrt{N+4}}\,,\qquad\qquad \lambda^{(1)}=\frac{4-N+\sqrt{N^2+152N+16}}{20(N+4)}\,.
\end{equation}
Plugging this into the expression for $A$, and using the form of the matrices $d_1$ and $d_2$ to evaluate all of the traces, one finds that
\begin{equation}
\begin{split}
    \frac{\pi}{288}\lsp A_{\text{NJLY}}=&-\frac{N\pi}{48(N+4)}\varepsilon^2-\frac{1}{14400
   (N+4)^3}  \bigg((N^2+152N+16)\sqrt{N^2+152N+16}\\&+2 N \left(25( \nc{g}{1}{9} +6\nc{h}{0}{2})(N+4)^2-2 N (44 N+57)-1944\right)+64\bigg)\pi\varepsilon^3+\text{O}(\varepsilon^4)\,,
\end{split}
\end{equation}
which indeed correctly matches with $\widetilde{F}_{\text{NJLY}}$ for the choice $\nc{h}{0}{2}=\tfrac{1}{12}(13-2\nc{g}{1}{9})$.

One notices that there are no unfixed degrees of freedom in the $\text{O}(\varepsilon^2)$ part of $A$ at either of these fixed points, and thus that it must exactly match the $\text{O}(\varepsilon^2)$ part of $\widetilde{F}$ off the bat. In fact, this agreement is highly non-trivial, and requires that the $\delta_{\circ\star}$ parts of the metric differ by an overall factor of $24$. Were (\ref{eq:g00fixing}) not to hold, it would not be possible to get $A$ to match $\widetilde{F}$ at both scalar and scalar-fermion fixed points simultaneously. As $\nc{a}{2}{1}\sim\text{O}(\varepsilon)$, the only terms contributing to the $\text{O}(\varepsilon^2)$ part of $A$ will be
\begin{equation}
    A^{(2)}=\nc{g}{0}{}\big(-\tfrac{1}{4}\varepsilon\Tr(y_i y_i)+\tfrac{1}{4}\Tr(y_i y_i y_j y_j)+\tfrac{1}{2}\Tr(y_i y_j y_i y_j)+\tfrac{1}{8}\Tr(y_i y_j)\Tr(y_i y_j)\big)\,,
\end{equation}
where we have plugged in the solution from (\ref{eq:01loopsol}) and the (scheme-invariant) values of the beta function coefficients. Using the fact that $y_{iab}$ will lie at a fixed point of the one-loop beta function, we can rewrite this as
\begin{equation}\label{eq:A1}
    A^{(2)}=-\tfrac{1}{8}\nc{g}{0}{}\varepsilon \Tr(y_i y_i)\,.
\end{equation}
Let us now compare this to the value of $\widetilde{F}$ at the same order in $\varepsilon$, using the treatment of $\widetilde{F}$ in \cite{Fei:2016sgs}. As the factor of $\sin\left(\frac{d\pi}{2}\right)\sim\varepsilon$, the only vacuum diagram which can contribute to $\widetilde{F}$ at $\text{O}(\varepsilon^2)$ is the one containing two $y$-vertices,
\begin{equation}
    \begin{tikzpicture}[scale=0.5,baseline=(vert_cent.base),square/.style={regular polygon,regular polygon sides=4}]
        \node (vert_cent) at (0,0) {$\phantom{\cdot}$};
        \draw[dashed] (-1,0) arc (180:0:1cm);
        \draw[thick] (-1,0) arc (180:360:1cm);
        \draw[thick] (-1,0) -- (1,0);
        \node[inner sep=0pt,outer sep=0pt,square, draw,fill=white,scale=3pt] at (-1,0) {};
        \node[inner sep=0pt,outer sep=0pt,square, draw,fill=white,scale=3pt] at (1,0) {};
    \end{tikzpicture}\,.
\end{equation}
Following \cite{Fei:2016sgs}, the value of this diagram can be computed on the sphere and can be written as $I_2\left(\frac{3d}{2}-2\right)$, using the family of integrals\cite{Fei:2015oha}
\begin{equation}
    I_2(\Delta)=\int\frac{d^{\llsp d}x d^{\llsp d}y\sqrt{g_x}\sqrt{g_y}}{s(x,y)^{2\Delta}}=(2R)^{2(d-\Delta)}\frac{2^{1-d}\pi^{d+\frac{1}{2}}\Gamma\left(\frac{d}{2}-\Delta\right)}{\Gamma\left(\frac{d+1}{2}\right)\Gamma(d-\Delta)}\,,
\end{equation}
where $s(x,y)$ is the chordal distance on the sphere, and $x$ and $y$ are stereographic coordinates. The value of $\widetilde{F}$ at any fixed point will thus be through $\text{O}(\varepsilon^2)$
\begin{equation}\label{eq:F1}
\widetilde{F}^{(2)}=\frac{\pi}{2}\left(-\frac{(4\pi)^2}{2}\right)\bigg(\lim_{d\rightarrow4} C_\phi C_\psi^2\lsp I_2\left(\frac{3d}{2}-2\right)\bigg)\Tr(y_i y_i)\varepsilon=-\frac{\pi}{96}\Tr(y_iy_i)\varepsilon\,,
\end{equation}
where we note that there is an extra factor of $16\pi^2$ relative to \cite{Fei:2016sgs} which arises because we have rescaled the couplings to remove that factors from the fixed point values (\ref{eq:GNYcrit}) and (\ref{eq:NJLYcrit}).

However, if $y_{iab}=0$ at the fixed point, i.e.\ for multiscalar models, then the leading term in both $A$ and $\widetilde{F}$ will instead be $\text{O}(\varepsilon^3)$. For these purely scalar models, the $\text{O}(\varepsilon^3)$ part of $A$ will be
\begin{equation}
    A^{(3)}=-\tfrac{1}{2}\varepsilon\lambda_{ijkl}\lambda_{ijkl}+\lambda_{ijmn}\lambda_{mnkl}\lambda_{ijkl}\,.
\end{equation}
For critical $\lambda_{ijkl}$, this can be re-writen as
\begin{equation}\label{eq:A2}
    A^{(3)}=-\tfrac{1}{6}\varepsilon\lambda_{ijkl}\lambda_{ijkl}\,.
\end{equation}
$\widetilde{F}$, on the other hand, will recieve a single contribution, from the diagram
\begin{equation}
    \begin{tikzpicture}[scale=0.5,baseline=(vert_cent.base)]
        \node (vert_cent) at (0,0) {$\phantom{\cdot}$};
        \draw[dashed] (0,0) circle (1cm);
        \draw[dashed] (1,0) to[out=135, in=45] (-1,0);
        \draw[dashed] (1,0) to[out=225, in=315] (-1,0);
        \node[inner sep=0pt,outer sep=0pt,circle, draw,fill=black,scale=3pt] at (-1,0) {};
        \node[inner sep=0pt,outer sep=0pt,circle, draw,fill=black,scale=3pt] at (1,0) {};
    \end{tikzpicture}\,.
\end{equation}
This diagram can easily be computed on the sphere \cite{Fei:2015oha} to find
\begin{equation}\label{eq:F2}
    \widetilde{F}^{(3)}=\frac{\pi}{2}\left(-\frac{(16\pi^2)^2}{2\cdot 4!}\right)\left(\lim_{d\rightarrow4}C_\phi^4\lsp I_2(2d-4)\right)=-\frac{\pi}{1728}(\lambda_{ijkl}\lambda_{ijkl})\varepsilon\,.
\end{equation}
If we are to cause $A$ to match $\widetilde{F}$ by rescaling, then the ratio of (\ref{eq:F1}) to (\ref{eq:F2}) must be equal to the ratio of (\ref{eq:A1}) to (\ref{eq:A2}). This is true only if
\begin{equation}
    \nc{g}{0}{}=24\,,
\end{equation}
which one can take to be a constraint imposing that $\widetilde{F}$ must appear in the solution space of $A$. Examining the preceding argument, one sees that the particular values of the one-loop beta function coefficients do not affect this constraint in any way, and thus that it is of an essentially different type to the constraints on the beta function coefficients following from the gradient flow equation. The fact that this constraint is satisfied by the value of $\nc{g}{0}{}$ demanded by the gradient flow equation at 1/2 loop order, given in (\ref{eq:g00fixing}), is highly non-trivial and suggests that $\beta^\star$ must contain information about $\widetilde{F}$. This lends credence to the conjecture that $\widetilde{F}$ can be consistently extended away from the CFTs to produce a gradient monotone.

\section{Numerics for fixed points}\label{sec:numerics}
In this section, we present the results of numerical searches for solutions of the one-loop beta functions for $1\leq N_s \leq4$ and $1 \leq N_f \leq2$. We find that the proportion of fixed points with non-zero beta shift increases rapidly as the numbers of fields increases. In particular for $N_s=4$, $N_f=2$, the vast majority of fixed points with non-zero Yukawa couplings have non-zero beta shift. These results demonstrate that the beta shift is generically non-zero for fixed points of theories with high numbers of fields and relatively small symmetry groups.

Table \ref{tab:propdata} shows the results of these numerical searches, including the proportion of fixed point with non-zero beta shift, for varying numbers of scalars and fermions. We solved for Dirac fermions in separate runs. Although the case of Dirac fermions is included in that of Weyl fermions, organisation in terms of Dirac fermions imposes additional constraints on the allowed Yukawa couplings. For instance, the Lagrangian \eqref{eq:Lagtwotwo} for $N_s=2$ and $N_f=2$ has six complex Yukawa couplings, whereas when the two Weyl fermions are combined into one Dirac fermion, the corresponding Lagrangian only has four real couplings. Thus, theories with Dirac fermions are a subspace of the theories with Weyl fermions of significantly lower dimension. Because of this, the numerical solver is less likely to find fixed points for which the organisation into Dirac fermions happens automatically, and one needs to impose the constraints by hand in order to find such fixed points. Note that $N_{f_D}$ denotes the number of Dirac fermions.
\begin{table}[H]
    \centering
    \caption{Fixed points of the one-loop beta functions found numerically}\label{tab:propdata}
    \begin{tabular}{|l|c|c|c|}
        \hline
        \textbf{Field content} & \textbf{FPs} & \textbf{$S \neq 0$} & \textbf{\% $S \neq 0$} \\
        \hline\hline
        $N_s = 2, N_f = 1$ & 13 & 1 & 8 \\
        \hline
        $N_s=2,N_f=2$ & $31$ & $4$ & $13$ \\
        \hline
        $N_s=2, N_{f_D}=1$&11&3&27\\
        \hline
        $N_s=2,N_{f_D}=2$&32&4&13\\
        \hline
        $N_s=3,N_f=1$ & $44$ & $16$ & $36$ \\
        \hline
        $N_s=3,N_f=2$ & $403$ & $276$ & $68$ \\
        \hline
        $N_s=4,N_f=1$ & $75$ & $38$ & $51$ \\
        \hline
        $N_s=4,N_f=2$ & $294$ & $190$ & $65$ \\
        \hline
    \end{tabular}
\end{table}

In Figs \ref{fig:2,1}--\ref{fig:D2,2} we have plotted the results of the numerics. For these plots we have used two invariants, $R$ and $T'$. These invariants satisfy a parabolic bound \cite{Pannell:2023tzc},
\begin{equation}
    R^2 + T' \leq \tfrac18 N_s\,.
\end{equation}
The relevant parabola has been plotted in each figure, demonstrating that the fixed points satisfy the bound in all cases. Fixed points with non-zero beta shift are plotted as red diamonds, while those with zero beta shifted are plotted as green dots.

\begin{figure}[H]
\centering
\begin{tikzpicture}
\begin{axis}[
    xmin=-1, xmax=1,
    ymin=-0.2, ymax=0.3,
    xlabel= $R$,
    ylabel= $T'$,
    ylabel style={rotate=-90},
    title={Scalar-fermion fixed points for $N_s=2$ and $N_f=1$},
    legend pos = outer north east
    ]
 \addplot+[
    only marks,
    mark=*,
    mark options={color=Dark2-A},
    mark size=2pt
    ]
table[x=r,y=t]{RT_Plots/2,1.dat};
 \addplot+[
    only marks,
    mark=diamond*,
    mark options={color=Dark2-B},
    mark size=4pt
    ]
table[x=r,y=t]{RT_Plots/2,1_S.dat};

\addplot[domain=-2:2,
    samples=1000,
    color=Dark2-C]{-x^2 + 2/8};

\end{axis}
\end{tikzpicture}
\caption{Results of numerical search for $N_s=2$, $N_f=1$. We find 13 fixed points, with 1 out of 13, or $\sim8\%$, having non-zero beta shift.}
\label{fig:2,1}
\end{figure}

\begin{figure}[H]
\centering
\begin{tikzpicture}
    \begin{axis}[
        xmin=-1, xmax=1,
        ymin=-0.2, ymax=0.3,
        xlabel= $R$,
        ylabel= $T'$,
        ylabel style={rotate=-90},
        title={Scalar-fermion fixed points for $N_s=2$ and $N_f=2$},
        legend pos = outer north east
        ]
     \addplot+[
        only marks,
        mark=*,
        mark options={color=Dark2-A},
        mark size=2pt
        ]
    table[x=r,y=t]{RT_Plots/2,2.dat};
     \addplot+[
        only marks,
        mark=diamond*,
        mark options={color=Dark2-B},
        mark size=4pt
        ]
    table[x=r,y=t]{RT_Plots/2,2_S.dat};
    \addplot[domain=-2:2,
        samples=1000,
        color=Dark2-C]{-x^2 + 2/8};
    \end{axis}
\end{tikzpicture}
\caption{Results of numerical search for $N_s=2$, $N_f=2$. We find 31 fixed points, with 4 out of the 31, or $\sim13\%$, having non-zero beta shift.}
\label{fig:2,2}
\end{figure}

\begin{figure}[H]
\centering
\begin{tikzpicture}
\begin{axis}[
    xmin=-1, xmax=1,
    ymin=-0.2, ymax=0.3,
    xlabel= $R$,
    ylabel= $T'$,
    ylabel style={rotate=-90},
title={Scalar-fermion fixed points for $N_s=2$ and $N_{f_D}=1$},
    legend pos = outer north east
    ]
 \addplot+[
    only marks,
    mark=*,
   mark options={color=Dark2-A},
    mark size=2pt
    ]
table[x=r,y=t]{RT_Plots/D2,1.dat};
 \addplot+[
    only marks,
    mark=diamond*,
   mark options={color=Dark2-B},
    mark size=4pt
    ]
table[x=r,y=t]{RT_Plots/D2,1_S.dat};

\addplot[domain=-2:2,
    samples=1000,
    color=Dark2-C]{-x^2 + 2/8};

\end{axis}
\end{tikzpicture}
    \caption{Results of numerical search for $N_s=2$, $N_{f_D}=1$. We find 11 fixed points, with 3 out of the 11, or $\sim27\%$ having non-zero beta shift.}
\label{fig:D2,1}
\end{figure}

\begin{figure}[H]
\centering
\begin{tikzpicture}
\begin{axis}[
    xmin=-1, xmax=1,
    ymin=-0.2, ymax=0.5,
    xlabel= $R$,
    ylabel= $T'$,
    ylabel style={rotate=-90},
    title={Scalar-fermion fixed points for $N_s=3$ and $N_f=1$},
    legend pos = outer north east
    ]
 \addplot+[
    only marks,
    mark=*,
   mark options={color=Dark2-A},
    mark size=2pt
    ]
table[x=r,y=t]{RT_Plots/3,1.dat};
 \addplot+[
    only marks,
    mark=diamond*,
   mark options={color=Dark2-B},
    mark size=3pt
    ]
table[x=r,y=t]{RT_Plots/3,1_S.dat};
\addplot[domain=-2:2,
    samples=1000,
    color=Dark2-C]{-x^2 + 3/8};
\end{axis}
\end{tikzpicture}
    \caption{Results of numerical search for $N_s=3$, $N_f=1$. We find 44 fixed points, with 16 out of the 44, or $\sim36\%$ having non-zero beta shift.}
\label{fig:3,1}
\end{figure}

\begin{figure}[H]
\centering
\begin{tikzpicture}
\begin{axis}[
    xmin=-1, xmax=1,
    ymin=-0.2, ymax=0.5,
    xlabel= $R$,
    ylabel= $T'$,
    ylabel style={rotate=-90},
    title={Scalar-fermion fixed points for $N_s=3$ and $N_f=2$},
    legend pos = outer north east
    ]
 \addplot+[
    only marks,
    mark=*,
   mark options={color=Dark2-A},
    mark size=2pt
    ]
table[x=r,y=t]{RT_Plots/3,2.dat};
 \addplot+[
    only marks,
    mark=diamond*,
   mark options={color=Dark2-B},
    mark size=3pt
    ]
table[x=r,y=t]{RT_Plots/3,2_S.dat};
\addplot[domain=-2:2,
    samples=1000,
    color=Dark2-C]{-x^2 + 3/8};
\end{axis}
\end{tikzpicture}
    \caption{Results of numerical search for $N_s=3$, $N_f=2$. We find 403 fixed points, with 276 out of the 403, or $\sim68\%$ having non-zero beta shift.}
\label{fig:3,2}
\end{figure}

\begin{figure}[H]
\centering
\begin{tikzpicture}
\begin{axis}[
    xmin=-1, xmax=1,
    ymin=-0.2, ymax=0.6,
    xlabel= $R$,
    ylabel= $T'$,
    ylabel style={rotate=-90},
    title={Scalar-fermion fixed points for $N_s=4$ and $N_f=1$},
    legend pos = outer north east
    ]
    \addplot+[
    only marks,
    mark=*,
    mark options={color=Dark2-A},
    mark size=2pt
    ]
    table[x=r,y=t]{RT_Plots/4,1.dat};
    \addplot+[
    only marks,
    mark=diamond*,
    mark options={color=Dark2-B},
    mark size=3pt
    ]
    table[x=r,y=t]{RT_Plots/4,1_S.dat};
    \addplot[domain=-2:2,
    samples=1000,
    color=Dark2-C]{-x^2 + 4/8};
\end{axis}
\end{tikzpicture}
\caption{Results of numerical search for $N_s=4$, $N_f=1$. We find 75 fixed points, with 38 out of the 75, or $\sim51\%$ having non-zero beta shift.}
\label{fig:4,1}
\end{figure}

\begin{figure}[H]
\centering
\begin{tikzpicture}
\begin{axis}[
    xmin=-1.5, xmax=1.5,
    ymin=-1, ymax=0.6,
    xlabel= $R$,
    ylabel= $T'$,
    ylabel style={rotate=-90},
    title={Scalar-fermion fixed points for $N_s=4$ and $N_f=2$},
    legend pos = outer north east
    ]
 \addplot+[
    only marks,
    mark=*,
    mark options={color=Dark2-A},
    mark size=2pt
    ]
table[x=r,y=t]{RT_Plots/4,2.dat};
 \addplot+[
    only marks,
    mark=diamond*,
    mark options={color=Dark2-B},
    mark size=3pt
    ]
table[x=r,y=t]{RT_Plots/4,2_S.dat};

\addplot[domain=-2:2,
    samples=1000,
    color=Dark2-C]{-x^2 + 4/8};
\end{axis}
\end{tikzpicture}
\caption{Results of numerical search for $N_s=4$, $N_f=2$. We find 294 fixed points, with 190 out of the 294, or $\sim65\%$ having non-zero beta shift.}
\label{fig:4,2}
\end{figure}

\begin{figure}[H]
\centering
\begin{tikzpicture}
\begin{axis}[
    xmin=-1.5, xmax=1.5,
    ymin=-1, ymax=0.5,
    xlabel= $R$,
    ylabel= $T'$,
    ylabel style={rotate=-90},
    title={Scalar-fermion fixed points for $N_s=2$ and $N_{f_D}=2$},
    legend pos = outer north east
    ]
 \addplot+[
    only marks,
    mark=*,
    mark options={color=Dark2-A},
    mark size=2pt
    ]
table[x=r,y=t]{RT_Plots/D2,2.dat};
 \addplot+[
    only marks,
    mark=diamond*,
    mark options={color=Dark2-B},
    mark size=3pt
    ]
table[x=r,y=t]{RT_Plots/D2,2_S.dat};

\addplot[domain=-2:2,
    samples=1000,
    color=Dark2-C]{-x^2 + 2/8};
\end{axis}
\end{tikzpicture}
\caption{Results of numerical search for $N_s=2$, $N_{f_D}=2$. We find 32 fixed points, with 4 out of the 32, or $\sim13\%$ having non-zero beta shift.}
\label{fig:D2,2}
\end{figure}

\section{Conclusion}\label{sec:conc}
The results of this work provide further evidence for the gradient structure of the renormalisation group. The non-trivial nature of the gradient conditions we get, which are satisfied whenever the full set of results needed to check them is available, indicates that the gradient property is a fundamental aspect of the perturbative RG. The proper framing of the gradient question necessarily involves the beta shift, which alters the trace of the energy-momentum tensor \cite{Jack:1990eb, Osborn:1991gm}, and we have also discussed the importance of the beta shift for questions of scale without conformal invariance and the precise definition of the dimension of operators at a conformal fixed point.

Our results do not imply anything about the gradient property of the RG at higher loop orders, and an argument valid to all orders in perturbation theory would be of great importance. It was illustrated in \cite{Jack:1990eb, Osborn:1991gm} that if one relaxes the Riemannian requirement on the metric, then the essential relation
\begin{equation}\label{eq:gradflowT}
    \frac{\partial A}{\partial g^I}=T_{IJ}B^J
\end{equation}
holds in $d=4$. The results of our work show that there exists a scheme in which $T_{IJ}$ of \eqref{eq:gradflowT} can be chosen to be symmetric, to the order checked. This is also known to be the case even if gauge fields are present \cite{Jack:2014pua, Poole:2019kcm}, at least at the loop orders checked in those works. A proof that this can be arranged to all loop orders would amount to a proof of the gradient property of the RG to all loops in $d=4$. The extension of this to $d=4-\varepsilon$ and its relevance to the strongly coupled physics of $d=3$ would likely require some new ideas.

\ack{We thank J.-F.\ Fortin for discussions and collaboration in the initial stages of the project. We also thank T.\ Steudtner for sharing with us his results \cite{Steudtner:2025blh} before publication. We are especially grateful to H.\ Osborn for enlightening discussions and comments on a draft of this manuscript. AS is supported by the Royal Society under grant URF\textbackslash{}R1\textbackslash211417 and by STFC under grant ST/X000753/1.}

\begin{appendices}

\section{One-loop beta functions and coupling invariants}\label{app:betasinvs}
In this appendix, following \cite{Pannell:2023tzc}, we discuss the invariants of the couplings used to distinguish the fixed points found numerically. Recall from \eqref{eq:Lint} that the theory is defined by one real symmetric tensor $\lambda_{ijkl}$ and one complex tensor $y_{iab}$, which is symmetric in $a$ and $b$. We have $i=1,\dots N_s$, $a=1,\dots N_f$, where $N_s$ and $N_f$ are the numbers of scalars and (Weyl) fermions respectively. We will sometimes write $y_i$ and omit the fermion flavour indices $a,b$. Expressions like $y_iy^*{\!\!\!}_j$ are then to be understood as matrix multiplication over the $ab$ indices, similarly $\Tr(y_iy^*{\!\!\!}_j)$ refers to the trace over those indices. The one-loop beta functions in terms of these tensors are given by
\begin{align}
    \beta^{\lambda}_{ijkl}&=-\varepsilon\lsp\lambda_{ijkl}+S_{3,ijkl}\lsp\lambda_{ijmn}\lambda_{mnkl}-4\lsp S_{6,ijkl}\lsp y_{iab}\lsp y^*{\!\!\!}_{jbc}\lsp y_{kcd}\lsp y^*{\!\!\!}_{lda}\nonumber\\
    &\hspace{6cm}+\tfrac{1}{2}S_{4,ijkl}\lsp(y_{iab}\lsp y^*{\!\!\!}_{mba}+y_{mab}\lsp y^*{\!\!\!}_{iba})\lambda_{mjkl}\,,\label{eq:betalambdaweyl}\\
    \beta^{y}_{iab}&=-\tfrac{1}{2}\varepsilon\lsp y_{iab}+\tfrac{1}{2}(y_{jac}\lsp y^*{\!\!\!}_{jcd}\lsp y_{idb}+y_{iac}\lsp y^*{\!\!\!}_{jcd}\lsp y_{jdb})+2\lsp y_{jac}\lsp y^*{\hspace{-5pt}}_{icd}\lsp y_{jdb}+\tfrac12(y_{icd}\lsp y^*{\!\!\!}_{jdc}+y_{jcd}\lsp y^*{\hspace{-5pt}}_{idc})\lsp y_{jab}\,,\label{eq:betayweyl}
\end{align}
where $S_{n,ijkl}$ indicates the sum over the $n$ permutations of the indices $i, j, k, l$ needed to make the expression symmetric. Solutions to these equations belong to $O(N_s)\times U(N_f)$ orbits. Transforming the couplings in this manner simply amounts to a field redefinition and thus solutions related by an $O(N_s)\times U(N_f)$ transformation are equivalent.

To properly distinguish fixed points we define invariants of the coupling tensors under $O(N_s)\times U(N_f)$ \cite{Pannell:2023tzc}. First we define the $O(N_s)$ tensors
\begin{equation}
    \begin{aligned}
        Z_{ij}&=\big(y_{iab}\lsp y^{*}{\hspace{-5pt}}_{jba}+y^{*}{\hspace{-5pt}}_{iab}\lsp y_{jba}\big)=\Tr(y_iy^*{\!\!\!}_j+y^*{\!\!\!}_iy_j)\,,\\
        X_{ijkl}&=\tfrac{1}{2}\lsp S_{6,ijkl}\lsp y_{iab}\lsp y^{*}{\hspace{-5pt}}_{jbc}\lsp y_{kcd}\lsp y^{*}{\hspace{-5pt}}_{lda}=\tfrac{1}{2}\lsp S_{6,ijkl}\Tr(y_{i}\lsp y^{*}{\hspace{-5pt}}_{j}\lsp y_{k}\lsp y^{*}{\hspace{-5pt}}_{l}\lsp)\,.
    \label{eq:zweyl}
    \end{aligned}
\end{equation}
It is useful to decompose the $O(N_s)$ tensors into irreducible components,
\begin{equation}
    \begin{aligned}
        Z_{ij}&=Z_0\lsp\delta_{ij}+Z_{2,ij}\,, \\
        X_{ijkl}&=X_0\lsp S_{3,ijkl}\lsp \delta_{ij}\delta_{kl}+S_{6,ijlk}\lsp X_{2,ij}\lsp \delta_{kl}+X_{4,ijkl}\,,\\
        \lambda_{ijkl}&=d_0\lsp S_{3,ijkl}\lsp\delta_{ij}\delta_{kl}+S_{6,ijlk}\lsp d_{2,ij}\delta_{kl}+d_{4,ijkl}\,,
    \label{eq:tensordecomposition}
    \end{aligned}
\end{equation}
where $Z_2,X_2,X_4,d_2$ and $d_4$ are all traceless symmetric tensors. Using these we can define the invariants
\begin{equation}
    \begin{aligned}
        a_0&=N_s(N_s+2)d_0=\lambda_{iijj}\,,\\
        a_1&=\lambda_{iimn}\lambda_{jjmn}\,,\\
        a_2&=||(N_s+4)d_2||^2=a_1-\frac{1}{N_s}a_0^2\,,\\
        S&=\lambda_{ijkl}\lambda_{ijkl}=||\lambda||^2\,,\\
        b_0&=Z_{ii}=N_sZ_0\,,\qquad \tilde{b}_0=a_0Z_0\,,\\
        b_1&=||Z_2||^2\,,\\
        b_2&=||(N_s+4)d_2+Z_2||^2=a_2+b_1+2(N_s+4)d_2\cdot Z_2\,,\\
        b_3&=X_{iijj}=N_s(N_s+2)X_0\,,\\
        Y&=||y_iy^*{\hspace{-5pt}}_i\llsp ||^2=\Tr(y_{i}y^*{\hspace{-5pt}}_{i}y_{j}y^*{\!\!\lnsp}_{j})\,.
    \end{aligned}
    \label{eq:invariants}
\end{equation}
For the purposes of plotting the fixed points, we can further define the invariants
\begin{equation}
    T'=S-6\lsp Y\,,\qquad R=\frac{1}{\sqrt{2N_s}}\Big(a_0+b_0-\frac12N_s\Big)\,.
\end{equation}
For fixed points, these invariants satisfy the parabolic bound \cite{Pannell:2023tzc}
\begin{equation}
    R^2 + T'\leq\tfrac18 N_s\lsp\varepsilon^2\,.
\end{equation}

In order to solve the beta function equations numerically we used the \href{https://www.wolfram.com/mathematica/}{\texttt{Mathematica}} package \href{http://www.xact.es/}{\texttt{xAct}} to expand out \eqref{eq:betalambdaweyl}, \eqref{eq:betayweyl} into components, yielding a system of coupled polynomials. We then solved these using the \texttt{FindRoot} function. For each solution the list of invariants \eqref{eq:invariants} was calculated. Equivalent fixed points with the same values for all invariants were filtered out leaving us with one representative per unique fixed point.

\section{Four loop constraints on the beta shift}\label{app:SPcon}
For the sake of completeness, and to permit the consistency of the gradient flow equations to be checked once the beta shifts have been computed at four loops, we include explicitly a list of the 64 constraints on $S_{ij}$ and $P_{ab}$ which are equivalent to $B^\star$ satisfying the gradient flow equations at 3/4 loop order, as well as an explicit expression for $S_{ij}$ and $P_{ab}$ at four loops.

\begin{equation}
\begin{split}
    (\nc{S}{4}{})_{ij}=&\nc{S}{4}{1}\,
\,.
\end{split}
\end{equation}
The constraints, in $\overline{\text{MS}}$, on the coefficients in $S$ and $P$ are then
\begin{align}
    \nc{S}{4}{8}+4 \nc{S}{4}{20}&=0\,,\\4 \nc{S}{4}{22}&=\nc{S}{4}{23}\,,\\\nc{P}{4}{16}+\nc{P}{4}{17}+\tfrac{3}{4}&=4 \nc{P}{4}{15}\,,\\ \nc{P}{4}{19}&=0\,,\\\nc{P}{4}{20}&=-\tfrac{3}{8}\,,\\8 \nc{P}{4}{16}+8 \nc{P}{4}{21}+3&=32 \nc{P}{4}{23}\,,\\8 (4 \nc{P}{4}{15}+\nc{P}{4}{21})&=8 \nc{P}{4}{16}+32 \nc{P}{4}{26}+3\,,\\ 8 (8 \nc{P}{4}{32}+16 \nc{P}{4}{33}+\zeta (3))&=32
   \nc{P}{4}{24}+23\,,\\\nc{P}{4}{45}&=0\,,\\16 \nc{S}{4}{10}+4 \nc{S}{4}{7}+7&=4 (4 \nc{S}{4}{26}+16 \nc{S}{4}{5}+\zeta (3))\,,\\\nc{S}{4}{22}+\nc{S}{4}{24}&=0\,,\\\nc{P}{4}{47}&=0\,,\\6 (\nc{P}{4}{16}+16 (\nc{P}{4}{1}+\nc{P}{4}{44}))&=6 \nc{P}{4}{21}+24 \nc{P}{4}{29}+24 \nc{P}{4}{34}+24 \nc{P}{4}{3}+1\,,\\\nc{S}{4}{17}&=0\,,\\\nc{S}{4}{13}&=\nc{S}{4}{6}+\tfrac{15}{16}\,,\\120
   \nc{P}{4}{18}+480 \nc{P}{4}{46}+\pi ^4+65&=0\,,\\4 \nc{P}{4}{38}&=\nc{P}{4}{6}\,,\\\nc{P}{4}{29}+\nc{P}{4}{43}+\nc{P}{4}{3}+\tfrac{31}{96}&=\nc{P}{4}{15}+4 \nc{P}{4}{1}\,,\\32 \nc{P}{4}{25}+37&=8 (2 \nc{P}{4}{22}+\zeta (3))\,,\\ 48 (8 \nc{P}{4}{49}+2 \nc{P}{4}{3}+\zeta (3))&=96 \nc{P}{4}{10}+5\,,\\24 \nc{P}{4}{36}+13&=6 \nc{P}{4}{37}+3 \zeta (3)\,,\\6 (4
   \nc{S}{4}{18}+\nc{S}{4}{4})&=3 \zeta (3)+4\,,\\\nc{S}{4}{12}&=4 \nc{S}{4}{6}+\zeta (3)-\tfrac{5}{2}\,,\\\nc{P}{4}{39}&=\nc{P}{4}{36}+\tfrac{31}{96}\,,\\48 \nc{P}{4}{9}&=192 \nc{P}{4}{41}+17\,,\\\nc{S}{4}{8}&=2 (2 \nc{S}{4}{11}+8 \nc{S}{4}{19}-8 \nc{S}{4}{6}-5 \zeta (3)+6)\,,\\ 48 (\nc{P}{4}{21}+4 \nc{P}{4}{34}+16 \nc{P}{4}{50}+3 \zeta (3))&=192 \nc{P}{4}{22}+193\,,\\16
   \nc{S}{4}{14}&=4 \nc{S}{4}{3}+7\,,\\\tfrac{1}{16}\nc{P}{4}{16}+\nc{P}{4}{1}+\nc{P}{4}{48}+\tfrac{5 \zeta (3)}{16}&=\tfrac{1}{4}\nc{P}{4}{29}+\nc{P}{4}{36}+\tfrac{1}{4}\nc{P}{4}{3}+\tfrac{529}{768}\,,\\\nc{P}{4}{4}&=0\,,\\\nc{P}{4}{18}+4 \nc{P}{4}{31}&=\nc{P}{4}{5}\,,\\ 4 \nc{P}{4}{28}+\nc{P}{4}{6}&=0\,,\\120 \nc{P}{4}{7}+40&=\pi ^4\,,\\60 \nc{S}{4}{21}+\pi ^4+215&=240
   \nc{S}{4}{1}\,,\\ 120 \nc{P}{4}{5}+\pi ^4+215&=480 \nc{P}{4}{8}\,,\\ \nc{S}{4}{2}&=\tfrac{35+\pi ^4}{120}\,,\\48 \nc{S}{4}{9}&=12 \nc{S}{4}{3}+41\,,\\ 3 \nc{P}{4}{9}+12 \nc{P}{4}{30}+1&=3 \nc{P}{4}{27}\,,\\2 \nc{S}{4}{8}+32 \nc{S}{4}{15}+5&=8 (\nc{S}{4}{11}+\nc{S}{4}{4})\,,\\48 (\nc{P}{4}{10}+\nc{P}{4}{29}+4 \nc{P}{4}{35})&=12 \nc{P}{4}{16}+12 \nc{P}{4}{21}+48
   \nc{P}{4}{34}+17\,,\\\nc{S}{4}{10}+4 \nc{S}{4}{16}+\nc{S}{4}{25}+\tfrac{1}{2}\nc{S}{4}{7}&=4 \nc{S}{4}{5}+\tfrac{9}{16}\,,\\48 \nc{P}{4}{27}&=192 \nc{P}{4}{42}+1\,,\\\nc{P}{4}{40}+4 \nc{P}{4}{2}+\tfrac{9}{64}&=\tfrac{1}{8} (8 \nc{P}{4}{12}+2 \nc{P}{4}{13}+\zeta (3))\,,\\\nc{P}{4}{12}+\nc{P}{4}{14}+\tfrac{\zeta (3)}{8}&=4
   \nc{P}{4}{11}+\tfrac{1}{4}\nc{P}{4}{13}+\tfrac{59}{192}\,,\\ 8 \nc{P}{4}{52}&=3\,,\\\nc{P}{4}{53}&=0\,,\\\nc{P}{4}{54}&=0\,,\\4 \nc{S}{4}{31}&=5\,,\\4 \nc{S}{4}{28}+\nc{S}{4}{33}+\tfrac{121}{48}&=0\,,\\ 4 \nc{P}{4}{55}&=\nc{P}{4}{57}\,,\\\nc{P}{4}{58}&=-\tfrac{1}{3}\,,\\ \nc{S}{4}{34}&=\tfrac{55}{48}\,,\\\nc{P}{4}{56}&=\tfrac{43}{96}\,,\\\nc{P}{4}{55}+\nc{P}{4}{59}+\tfrac{13}{96}&=0\,,\\\nc{S}{4}{27}&=0\,,\\24
   \nc{S}{4}{29}+121&=96 \nc{S}{4}{35}\,,\\24 \nc{S}{4}{29}&=96 \nc{S}{4}{32}+17\,,\\16 (\nc{P}{4}{51}+24 \nc{P}{4}{60})&=9\,,\\ \nc{P}{4}{61}&=\tfrac16\,,\\2 \nc{S}{4}{28}+\nc{S}{4}{30}+\tfrac{121}{96}&=\tfrac{1}{4}\nc{S}{4}{29}\,,\\\nc{S}{4}{37}&=\tfrac{13}{16}\,,\\\nc{S}{4}{36}&=\tfrac{23}{32}\,,\\\nc{S}{4}{38}&=\tfrac{13}{96}\,.
\end{align}

\end{appendices}

\bibliography{main}

\end{document}